\documentclass[12pt,letter]{article}
\pdfoutput=1
\usepackage{graphicx, epsfig, color,cite}
\usepackage{amsmath,amsthm,amsfonts,geometry,etoolbox}

\usepackage{amssymb}
\usepackage{caption,subcaption,graphicx,appendix}
\usepackage[hidelinks]{hyperref}
\def\to{\rightarrow}

\def\bi{\begin{itemize}}
	\def\ei{\end{itemize}}
\def\te{\tilde e}
\def\ta{\tilde a}

\def\tchi{\tilde\chi}

\def\tu{\tilde u}

\def\tst{\tilde t}
\def\ttau{\tilde \tau}

\def\tnu{\tilde\nu}
\def\tell{\tilde\ell}

\def\tw{\widetilde\chi^{\pm}}

\def\tz{\widetilde\chi^0}
\def\alt{\lesssim}
\def\agt{\gtrsim}
\def\be{\begin{equation}}  
	\def\ee{\end{equation}}  
\def\bea{\begin{eqnarray}}  
	\def\eea{\end{eqnarray}}

\newcommand{\myeq}{\begin{small}\begin{equation}\begin{aligned}}
			\newcommand{\myeqend}{\end{aligned}\end{equation}\end{small}}

\begin{document}
	\begin{titlepage}
		\begin{flushright}
			OU-HEP-240130
		\end{flushright}
		
		\vspace{0.5cm}
		\begin{center}
			{\Large \bf Supersymmetry with scalar sequestering}\\
			
			\vspace{1.2cm} \renewcommand{\thefootnote}{\fnsymbol{footnote}}
			{\large Howard Baer$^{1}$\footnote[1]{Email: baer@ou.edu },
				Vernon Barger$^2$\footnote[2]{Email: barger@pheno.wisc.edu},
				Dakotah Martinez$^1$\footnote[3]{Email: dakotah.s.martinez-1@ou.edu}
			}\\ 
			\vspace{1.2cm} \renewcommand{\thefootnote}{\arabic{footnote}}
			{\it 
				$^1$Homer L. Dodge Department of Physics and Astronomy,\\
				University of Oklahoma, Norman, OK 73019, USA \\[3pt]
			}
			{\it 
				$^2$Department of Physics,
				University of Wisconsin, Madison, WI 53706 USA \\[3pt]
			}
			
		\end{center}
		
		\vspace{0.5cm}
		\begin{abstract}
			\noindent
			Supersymmetric models with a strongly interacting superconformal
			hidden sector (HS) may drive soft SUSY breaking scalar masses,
			bilinear soft term $B\mu$ and Higgs combinations
			$m_{H_{u,d}}^2+\mu^2$ to small values at some intermediate scale,
			leading to unique sparticle mass spectra along with possibly
			diminished finetuning in spite of a large superpotential $\mu$ parameter.
			We set up a computer code to calculate such spectra,
			which are then susceptible to a variety of constraints:
			1. possible charge-or-color breaking (CCB)
			minima in the scalar potential,
			2. unbounded from below (UFB) scalar potential,
			3. improper electroweak symmetry breaking,
			4. a charged or sneutrino lightest SUSY particle (LSP),
			5. generating $m_h\sim 125$ GeV,
			6. consistency with LHC sparticle mass limits, and
			7. naturalness.
			We find this bevy of constraints leaves little or no viable parameter
			space for the case where hidden sector dynamics dominates MSSM running,
                        even for the case of non-universal gaugino masses.
			For the case with moderate HS running with comparable MSSM running,
                        and with universal gaugino masses,
                        then the finetuning is ameliorated, but nonetheless remains high.
			Viable spectra with moderate HS running and with low finetuning and
                        large $\mu$ can be found for non-universal gaugino masses.
		\end{abstract}
	\end{titlepage}
	
	\section{Introduction}
	\label{sec:intro}
	
	Particle physics models featuring weak scale supersymmetry\cite{Baer:2006rs,Dreiner:2023yus}
        (SUSY) are notable in that they contain a solution to the gauge hierarchy problem
        (GHP) and are
	supported by several virtual effects, including the celebrated unification of
	gauge couplings, a prediction for the Higgs mass within expectations
	from theory and experiment, and a prediction of a heavy top quark needed for radiative
	electroweak symmetry breaking (EWSB) well before the top quark was discovered.
	Even so, the non-appearance of supersymmetric matter at the CERN
	Large Hadron Collider\cite{Canepa:2019hph} (LHC) has potentially opened up
	a different naturalness problem\cite{Dine:2015xga}:
	the little hierarchy problem (LHP) concerning the burgeoning mass gap
	between the weak scale $m_{weak}\sim m_{W,Z,h}\sim 100$ GeV
	and the so-called soft SUSY breaking scale $m_{soft}\sim m_{sparticles}$,
	{\it i.e.} why is $m_{weak}\ll m_{soft}\agt 1-10$ TeV?
	
	A potential solution to the LHP comes from examining the explicit connection
	between $m_{weak}$ and $m_{soft}$ which arises from minimization of the
	Higgs (scalar) potential of the Minimal Supersymmetric Standard Model (MSSM):
	\be
	m_Z^2/2=\frac{(m_{H_d}^2+\Sigma_d^d)-(m_{H_u}^2+\Sigma_u^u)\tan^2\beta}{\tan^2\beta -1}-\mu^2\sim -m_{H_u}^2-\Sigma_u^u(\tst_{1,2})-\mu^2
	\label{eq:mzs}
	\ee
	where $m_{H_{u,d}}$ are the Higgs soft SUSY breaking masses, $\mu$ is the
	(SUSY conserving) Higgs mixing term, $\tan\beta=v_u/v_d$ is the ratio of
	Higgs field vevs and the $\Sigma_{u,d}^{u,d}$ terms contain an assortment of
	loop corrections (expressions may be found in Ref's \cite{Baer:2012cf}
	and \cite{Baer:2021tta}).
	Under {\it practical naturalness}\cite{Baer:2023cvi} --
	wherein all independent contributions to an observable ought to be comparable
	to or less than the observable --
	then only $|m_{H_u}|$ and $\mu$ ought to be $\sim m_{weak}$ whilst
	other sparticle contributions to the weak scale may be much heavier
	since their contributions are loop-suppressed.
	A naturalness measure $\Delta_{EW}$ has been proposed\cite{Baer:2012up}
	which compares the largest contribution to the right-hand-side of
	Eq. \ref{eq:mzs} to $m_Z^2/2$.
	Models with $\Delta_{EW}\alt 30$ are then presumed practically natural
	and for such models there is no LHP.
	
	One consequence of natural SUSY is that of all the sparticles,
	one expects only the higgsinos (with mass $\sim |\mu |$) to be of order $m_{weak}$.
	Higgsino pair production is difficult to observe
	at LHC due to the small inter-higgsino mass gap\cite{Baer:2011ec},
	and limits on $\mu$ vary between $\agt 100-200$ GeV depending on the mass gap
	$\Delta m_{21}=m_{\tchi_2^0}-m_{\tchi_1}^0$\cite{Baer:2020sgm}.
	A challenge for LHC Run 3 and high-lumi LHC (HL-LHC) is to either discover
	light higgsinos or else rule out natural SUSY by excluding the
	higgsino discovery plane\cite{Baer:2021srt}.
	At present, both ATLAS\cite{ATLAS:2019lng} and CMS\cite{CMS:2021edw}
	seem to have $\sim 2\sigma$ excesses
	in the opposite-sign-dilepton-plus-jet-plus-MET (OSDLJMET)
	signature\cite{Han:2014kaa,Baer:2014kya};
	upcoming new data should either confirm or exclude such signal channels.
	
	Unnatural SUSY with large $|\mu |\gg m_{weak}$, while possible, seems
	at first glance highly implausible. However, model builders have proposed
	a way to remain natural even with $|\mu |\gg m_{weak}$ by discovering models
	where the {\it combinations} $m_{H_{u,d}}^2+\mu^2$ are driven to be tiny,
	while $\sqrt{|m_{H_{u,d}}^2|}$ and $\mu$ individually can each be large at the
	weak scale.
	This method is called {\it scalar sequestering}
	(SS)\cite{Murayama:2007ge,Perez:2008ng,Kim:2009sy,Martin:2017vlf}.
	
	The method of hidden sector sequestering (HSS) of visible sector operators
	arises from postulating the existence of a strongly interacting
	nearly superconformal hidden sector (HS) which is operative between the
	messenger scale $M_*$ (taken to be of order the reduced Planck mass
	$\sim m_P$ in the case of gravity mediation) and a much lower intermediate
	scale $M_{int}$ where the superconformal symmetry
	is broken and SUSY is also broken.
        This method of sequestering was originally
	proposed\cite{Luty:2001zv} as a means to obtain anomaly-mediated SUSY breaking (AMSB)
	models\cite{Randall:1998uk,Giudice:1998xp} when
	geometric sequestering was shown to be difficult to realize\cite{Anisimov:2001zz}.
	
	Under HSS, the various soft SUSY breaking terms
	get squeezed to tiny values via RG running between $m_*$ and $M_{int}$
	by a power-law behavior:
	\be
	m_{soft}(M_{int})\sim (M_{int}/m_*)^\Gamma m_{soft}(m_*)
	\ee
	where the exponent $\Gamma $ includes combinations of classical and
	anomalous dimensions of HS fields $S$.
	$\Gamma$ is not directly calculable due to the strong dynamics in the HS
	but is instead  assumed to be $\sim 1$.
	For $M_{int}\sim 10^{11}$ GeV and $\Gamma\sim 1$, then the suppression of gravity-mediated
	soft terms can be $\sim 10^{-7}$ in which case the AMSB soft terms would
	be dominant.
	Additional symmetries seemed to be required in order
	for HSS to be viable; nonetheless, the lesson was that
	(model dependent) hidden sector effects could potentially modify the
	assumed running of SUSY model parameters as expected under the
	MSSM only\cite{Dine:2004dv,Craig:2009rk}.
	HSS was then found to offer a solution to the needed suppression
	of various problematic operators.
	For instance, in gauge mediation\cite{Giudice:1998bp} the $B\mu$ soft term
	is expected with $B\mu \gg \mu^2$, leading to the famous $B\mu/\mu$ problem.
	HSS could be used to suppress $B\mu (M_{int}) \sim 0$ thus solving the
	problem\cite{Roy:2007nz,Murayama:2007ge}.
	Also, in gravity mediation, scalar masses arise via hidden sector-visible
	sector couplings such as
	\be
	\int d^4\theta \frac{c_{ij}}{m_P^2}S^\dagger S Q_i^\dagger Q_j
	\label{eq:sseq}
	\ee
	where the $Q_i$ are visible sector fields and the $S$ are hidden sector fields
	which acquire an auxiliary field SUSY breaking vev $F_S\sim (10^{11}$ GeV)$^2$.
	In gravity-mediation, such operators are unsuppressed by any known symmetry
	(leading to the SUSY flavor problem), but could be squeezed to tiny
	values via scalar sequestering.
	A third application of (scalar) sequestering
	is to ameliorate the LHP while maintaining large $\mu$ values:
	$|\mu|\gg m_{weak}$. This case, which is the subject of the present paper, 
	makes use of Eq. \ref{eq:sseq} to suppress via hidden sector running all
	scalar masses to $\sim 0$. However, in the case where the
	Giudice-Masiero (GM) mechanism\cite{Giudice:1988yz} is
        assumed\footnote{Twenty solutions to the SUSY $\mu $
		problem are reviewed in Ref. \cite{Bae:2019dgg}} to generate a
	weak scale value of $\mu$,
	then the scalar sequestering actually applies to 
	$m_Q^2$ for matter scalars, but to the combinations
	$m_{H_{u,d}}^2+\mu^2$ for Higgs scalars.
	In this case, at the intermediate scale $M_{int}$, then one
	expects $m_Q^2\sim 0$ but with $m_{H_{u,d}}^2\sim -\mu^2$ so that
	$\mu$ can be large whilst the combination $m_{H_{u,d}}^2+\mu^2$ is small:
	this has the potential to fulfill the naturalness requirement in Eq. \ref{eq:mzs}
	while maintaining large $|\mu|\gg m_{weak}$ since $\mu^2$ and
	$m_{H_{u,d}}^2$ are no longer independent.
	
	In this paper, we examine the phenomenology of SUSY models with
	scalar sequestering.
	In Sec. \ref{sec:review}, we present a brief review of
	the theory underlying scalar sequestering. Two different
	theory approaches have emerged: strong scalar sequestering where hidden sector
	running overwhelms MSSM running\cite{Murayama:2007ge,Perez:2008ng},
	and moderate scalar sequestering\cite{Martin:2017vlf}, wherein
	hidden sector running and MSSM running are comparable, leading to quasi-fixed
	point behavior for the intermediate scale soft term boundary conditions.
	In Sec. \ref{sec:PRS}, we examine strong SS, dubbed here as the PRS
        (Perez, Roy and Schmaltz) scheme\cite{Perez:2008ng}.
	Here, the intermediate scale boundary conditions are so determinative
	that only one (or a few) parameters completely determine the SUSY phenomenology.
	In this case, problems emerge for appropriate electroweak symmetry
	breaking, vacuum stability, and dark matter physics
        (with typically a charged LSP and sometimes a left-sneutrino LSP).
	The latter case with a charged LSP can be dispensed with via either an
        assumed $R$-parity violation\cite{Barger:1989rk,Dreiner:1997uz}
	or assumed LSP decays to non-MSSM DM particles such as an axino $\ta$\cite{Choi:2011yf}.
        In Sec. \ref{sec:pspace}, we verify these results with parameter space scans in the
        PRS scheme with and without unified gaugino masses.
        In Sec. \ref{sec:SPM}, we instead adopt the scheme in \cite{Martin:2017vlf} -- we refer to this scheme as SPM (Stephen P. Martin)--
	with more limited HS running which is comparable to MSSM running.
	In this scheme, for the case of unified gaugino masses (UGM),
	we find that although SS reduces the amount of EW finetuning,
	significant weak scale finetuning
	arising from large top-squark masses remains, so that the finetuning problem
	cannot be said to be eliminated for large $\mu$.
	However, in the case of non-universal gaugino masses (NUGM) which lead
	to large stop mixing and $m_h\simeq 125$ GeV,
	then evidently low finetuning along with
	appropriate EWSB can be achieved for more moderate values of $\mu\sim 1$ TeV.
	Our findings are summarized in Sec. \ref{sec:conclude}.
	
	\section{Brief review of scalar sequestering}
	\label{sec:review}
	
	Let us assume a gravity-mediated generation of soft SUSY breaking terms
	since gauge-mediation gives rise to trilinear soft terms $A\sim 0$ and hence
	requires large, unnatural values of top squarks\cite{Baer:2014ica}
	to generate $m_h\sim 125$ GeV\cite{Carena:2002es}.
	At some scale $m_*<m_P$, the (superconformal) hidden sector becomes strongly interacting.
	Its coupling to visible sector fields leads to suppression of scalar soft breaking
	masses and also the bilinear soft term $b\equiv B\mu$.
	At some intermediate scale $M_{int}$, the conformal symmetry is broken and the
	hidden sector is integrated out of the low energy EFT. Also around this scale,
	SUSY is broken at a scale $Q_{SUSY}^2\sim F_S$.
	
	Under gravity-mediation, the following operators give rise to the usual
	soft terms:
	\be
	\int d^2\theta c_\lambda\frac{S}{m_P}WW +h.c. \Rightarrow m_\lambda\sim c_\lambda(F_S/m_P),
	\label{eq:inomass}
	\ee
	\be
	\int d^2\theta c_A\frac{S}{m_P}\phi_i\phi_j\phi_k +h.c. \Rightarrow A_{ijk}\sim c_A(F_s/m_P),
	\label{eq:Aterms}
	\ee
	\be
	\int d^4\theta c_{ij}\frac{R}{m_P^2}\phi_i^\dagger\phi_j\Rightarrow
	m_{\phi_{ij}}^2\sim c_{ij}(F_S/m_P)^2,
	\label{eq:mphi}
	\ee
	and
	\be
	\int d^4\theta c_b\frac{R}{m_P^2}H_uH_d +h.c.\Rightarrow B\mu\sim c_b(F_S/m_P)^2,
	\ee
	where $S$ is a HS chiral superfield and $R$ is a real product of
	hidden sector fields with $R\sim S^\dagger S +\cdots$.
	In addition, for the scalar sequestering model, one assumes the $\mu$
	term is initially suppressed (by some symmetry?) but then arises
	via the Giudice-Masiero\cite{Giudice:1988yz} mechanism at the scale $m_{soft}$ via
	\be
	\int d^4\theta c_\mu \frac{S^\dagger}{m_P}H_uH_d\Rightarrow \mu_{GM}\sim c_\mu (F_S/m_P) .
	\ee
	The holomorphic terms ($\int d^2\theta$) are protected against renormalization effects
	by non-renormalization theorems but the non-holomorphic terms are not.
	The latter terms give rise to scalar masses $m_{\phi_{ij}}^2$ and
	the bilinear soft term $B\mu$, and will scale
	between $m_*$ and $M_{int}$ as $(M_{int}/m_*)^{\Gamma}$ where
	the exponent $\Gamma$ is related to the anomalous dimension of the $S$ field.
	
	While $\Gamma$ is not directly calculable since the HS is strongly
	interacting, under the assumption that $\Gamma$ is large and positive,
	{\it e.g.} $\sim 1$, then the factor $(M_{int} /m_*)^{\Gamma}$ can lead
	to large suppression of scalar masses and $B\mu$ as compared to gaugino masses,
	$A$-terms and $\mu$.
	However, while $\mu$ can remain large under scalar sequestering,
	the combination $\hat{m}_{H_{u,d}}^2\equiv m_{H_{u,d}}^2+\mu^2$ gets driven
	to tiny values by the $(M_{int}/m_*)^{\Gamma}$ factor. 
	
	\section{Scalar sequestered SUSY: PRS boundary conditions}
	\label{sec:PRS}
	
	In the PRS scheme\cite{Murayama:2007ge,Perez:2008ng},
	the SS is assumed to dominate any MSSM running of soft terms.
	In this case, one expects the usual
	MSSM running for gaugino masses, $A$-terms and $\mu$ between the high scale
	$m_*$ and the intermediate scale $M_{int}$, whilst HS effects suppress
	matter scalar masses $m_{\phi_{ij}}^2$, $B\mu$ and Higgs combinations
	$m_{H_{u,d}}^2+\mu^2$.
	Thus, (under the assumption of unified gaugino masses)
	the parameter space of the model is given by
	\be
	m_{1/2},\ A_0,\ \mu\ \ \ {\rm and}\ \ \ M_{int}
	\ee
	where the first three of these are given at the high scale $m_*$.
	Motivated by gauge coupling unification, we take $m_*=m_{GUT}$, the scale
	where $g_1$ and $g_2$ unify under MSSM running,
	and where $m_{GUT}\simeq 2\times 10^{16}$ GeV.
	Meanwhile, the matter scalar masses, $B\mu$ and $m_{H_{u,d}}^2+\mu^2$
	are taken to be $\sim 0$ at the scale $Q\sim M_{int}$. 
	
	\subsection{Results for $M_{int} =10^{11}$ GeV and $A_0<0$}
	
\begin{figure}[!htbp]
		\begin{center}
			\includegraphics[height=0.4\textheight]{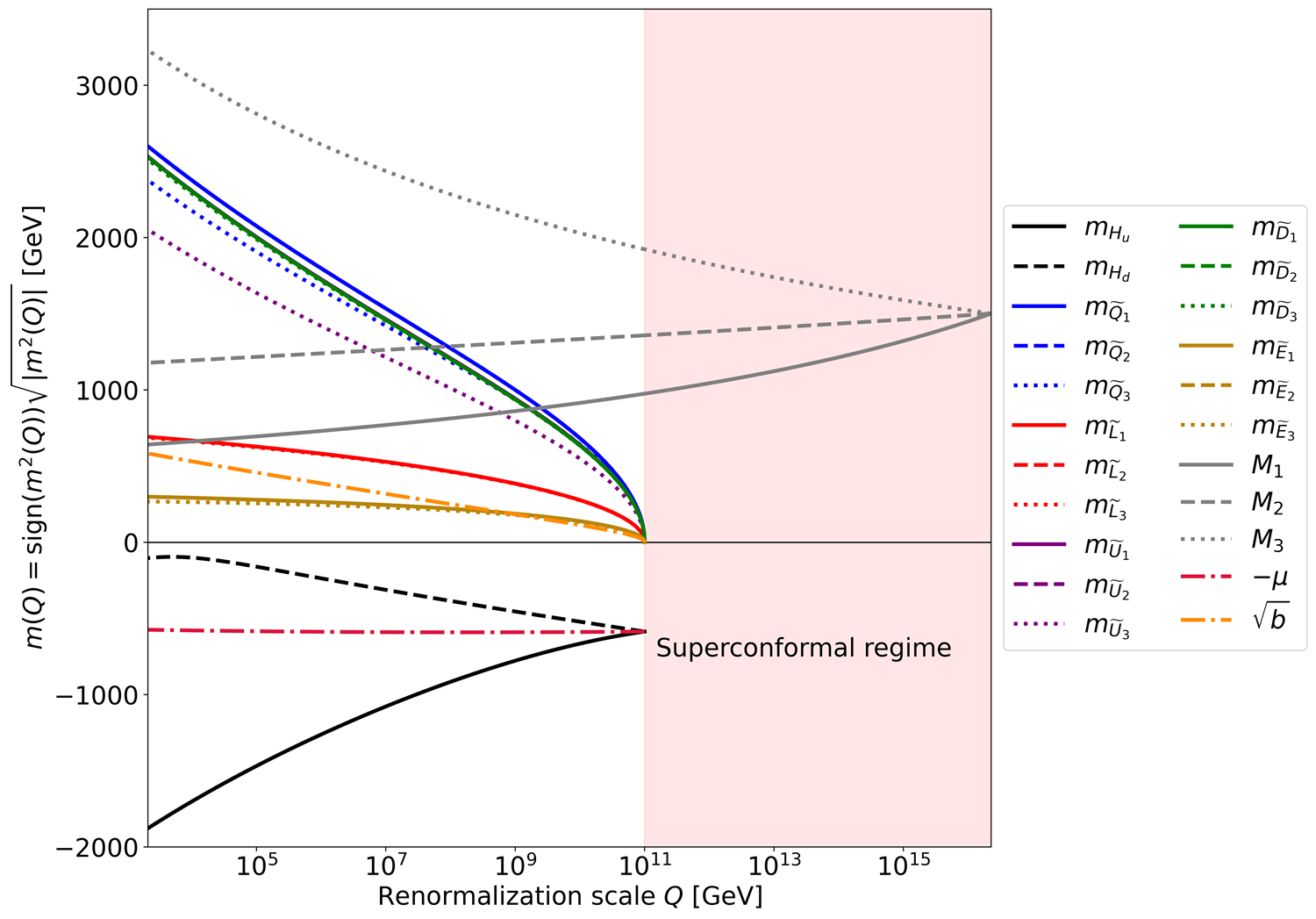}\\
			\includegraphics[height=0.4\textheight]{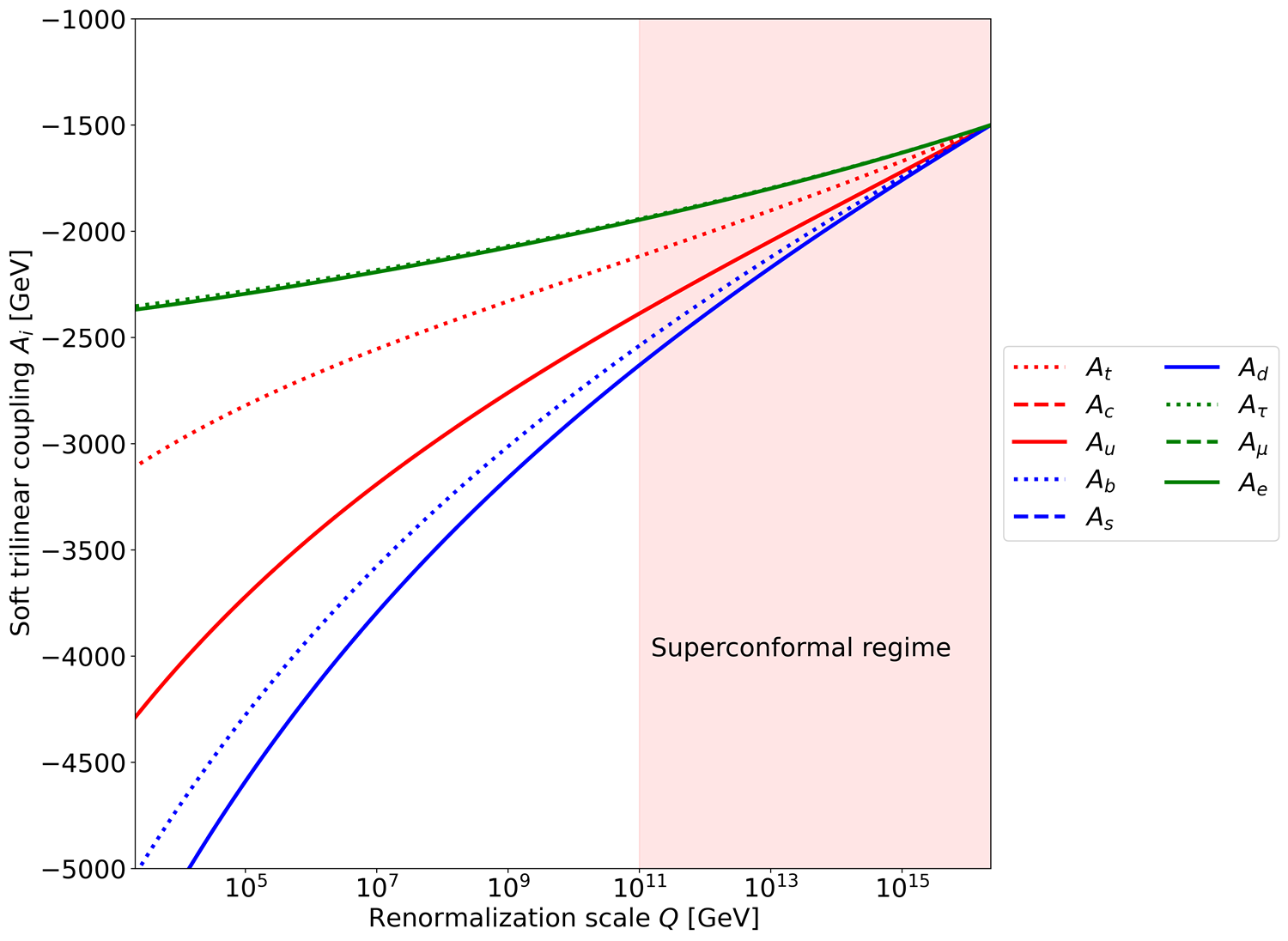}
			\caption{Running of soft terms and $-\mu$ in the PRS
				scalar sequestering scheme for $m_{1/2}=1.5$ TeV, $A_0=-m_{1/2}$,
				and $\mu = 500$ GeV. We also take the intermediate scale
				$M_{int} =10^{11}$ GeV. In frame {\it a}) we show running scalar masses
				and the $\mu$ term, while in frame {\it b}) we show the running trilinear
				soft terms.
				\label{fig:RGE11}}
		\end{center}
	\end{figure}

	As an illustration, we show in Fig. \ref{fig:RGE11} the running of soft terms
	and $\mu$ for the case where $m_{1/2}=-A_0=1.5$ TeV with $\mu =500$ GeV
	(the reason for $\mu \sim 500$ GeV is to be explained shortly).
	The pink shaded region shows the superconformal regime, whilst the soft terms
	run according to MSSM-only RGEs in the left-side unshaded region.
	We see from frame {\it a}) that the matter scalars start running at
	$Q=10^{11}$ GeV where the squark masses are pulled to large values
	$\agt 2$ TeV due to the influence of the $SU(3)$ gaugino mass $M_3$.
	Left-slepton masses are pulled up by a large $SU(2)_L$ gaugino mass $M_2$
	to the vicinity of $\sim 650$ GeV at $m_{weak}$ whilst the right slepton
	masses are pulled up by the $U(1)_Y$ gaugino mass $M_1$ to $\sim 300$ GeV. 
	The running of the bilinear $b$-term is given at the one-loop level by
	\be
	\frac{db}{dt}=\frac{\beta_b^{(1)}}{16\pi^2}
	\ee
	where the one-loop beta function is given by
	\be
	\beta_b^{(1)}=b(3f_t^2+3f_b^2+f_\tau^2-3g_2^2-\frac{3}{5}g_1^2)+\mu
	(6a_tf_t+6a_bf_b+2a_\tau f_\tau+6g2^2M_2+\frac{6}{5}g_1^2M_1)
	\label{eq:betab}
	\ee
	where the $f_i$ are Yukawa couplings, the $g_i$ are gauge couplings,
        the $a_i=A_if_i$ are the reduced trilinear couplings,
	and the $M_i$ are gaugino masses (further RGEs are given in, {\it e.g.},
	Ref. \cite{Baer:2006rs}, with their two-loop counterparts in Ref. \cite{Martin:1994rge}).
	In our numerical results presented in this paper, we use the full two-loop running of soft terms and gauge and Yukawa couplings. 
	
		The $\sqrt{b}=\sqrt{B\mu}$ term is pulled from zero at $Q=M_{int}$ to
	$\sim 550$ GeV at $Q= m_{weak}$ mainly by the second term of Eq. \ref{eq:betab}.
	Meanwhile, with $\mu =500$ GeV, the $\text{sign}(m_{H_{u,d}}^2)*\sqrt{|m_{H_{u,d}}|}$
	soft terms begin at $-500$ GeV and $m_{H_u}^2$ is driven to large negative
	values at $m_{weak}$ due to the large top-quark Yukawa coupling $f_t$.
	Also, $m_{H_d}$ is driven dominantly by the gaugino mass $M_2$ to small
	negative values $\sim -100$ GeV at $m_{weak}$.
	Frame {\it b}) shows the running of trilinear soft terms starting from
	$Q=m_{GUT}$. These terms are pushed to
	large negative values by the respective gauge interactions.
	In the case of $A_t$, this may help drive stop masses towards tachyonic
	values and consequently to charge and/or color breaking (CCB) minima
	in the scalar potential.

	A major check on this very constrained PRS scheme is if the EW symmetry is
	properly broken. Let us recall the (tree-level) conditions for proper EWSB.
	First, one must check whether the scalar potential
	indeed \emph{does not develop a minimum} at $h_u^0=h_d^0=0$,
	the origin of neutral scalar field space. The stability of the critical point satisfying
	\[
		\frac{\partial V}{\partial h_{u}^{0}}\Bigg|_{h_{u}^{0}=h_{d}^{0}=0}=\frac{\partial V}{\partial h_{d}^{0}}\Bigg|_{h_{u}^{0}=h_{d}^{0}=0}=0
	\]
	are determined by the nature of the eigenvalues of the matrix of second derivatives of the scalar potential, $V$, evaluated at the origin of field space. We refer to this matrix of second derivatives as the Hessian. Here, the neutral scalar fields are denoted $h_{u,d}^{0}$.
	
		The goal is to have a vacuum whose origin of field space is destabilized, else EWSB fails to occur properly.
	There are two cases in which this can happen:
	\begin{enumerate} 
		\item the origin is a maximum in field space, or perhaps
		\item the origin is a saddle point. 
	\end{enumerate}
	To determine the stability of the critical points we find, the type of critical point can be identified using the multivariate second partial derivative test.
	Case 1 occurs when the determinant of this Hessian is positive, but $m_{H_{u}}^{2}+\mu^{2}<0$ at the SUSY scale; then, the origin of field space will be a maximum.
	This secondary condition is crucial, meaning the positive determinant \emph{alone} is insufficient here to determine the nature of the critical point at the origin.
	When the determinant is positive, but $m_{H_{u}}^{2}+\mu^{2}>0$, then the origin of field space will be a \emph{minimum}, hence the scalar fields fail to acquire nonzero VEVs and EWSB fails to occur.
	
		Case 2 occurs when the determinant of the Hessian of the scalar potential at the origin
	with respect to the neutral Higgs scalars is negative, as this implies its eigenvalues have opposite signs,
	leading to
	\be
	(B\mu)^2>(m_{H_u}^2+\mu^2) (m_{H_d}^2+\mu^2) .
	\label{eq:8p17a}
	\ee
	This is often referred to in the literature as the condition for having a maximum at the origin of field space, but is more accurately described as a saddle point. In either case of a maximum or a saddle point, the origin is destabilized, so proper EWSB may yet be achievable, barring failure in the conditions below.
	In particular, given that $m_{H_u}^2$ is driven large negative and $B\mu$ is driven small
	positive, this saddle point condition may not always occur, but maxima sometimes occur instead as in case 1 and must be checked carefully!
	
		Secondly, one must check that the scalar potential is bounded from
	below (vacuum stability) in the $D$-flat direction $h_u^0=h_d^0$ leading
	to the requirement that
	\be
	m_{H_u}^2+m_{H_d}^2+2\mu^2 >2|B\mu |.
	\label{eq:8p17b}
	\ee
	Given that $m_{H_u}^2$ is large negative, this condition also may be subject to failure,
	in which case the scalar potential is unbounded from below (UFB).
	
	If an appropriate EWSB occurs, then minimization of the Higgs potential
	allows one to determine the Higgs vevs $v_u$ and $v_d$, with
	$\tan\beta =v_u/v_d$ as usual.
	The minimization conditions can be recast at tree-level as
	\be
	B\mu =\frac{\left(m_{H_u}^2+m_{H_d}^2+2\mu^2\right)\sin (2\beta)}{2}
	\label{eq:Bmu}
	\ee
	and
	\be
	m_Z^2/2=\frac{m_{H_d}^2-m_{H_u}^2\tan^2\beta}{\tan^2\beta -1}-\mu^2 .
	\label{eq:mzs_ewsb}
	\ee
	
	Usually, in models like mSUGRA,
	the first of these is used to trade $B\mu$ for $\tan\beta$ and the second
	is used to determine the magnitude of $\mu$. In the present case, since
	the boundary condition for $B\mu$ is $\sim 0$ at $Q=M_{int}$,
	it is not available to determine a unique value of $\tan\beta$, since the
	running of the soft parameters depends on the Yukawa couplings which in
	turn depend on $v_u$ and $v_d$, whose values then define $\tan\beta$.
	Furthermore, from Eq. \ref{eq:mzs_ewsb} we see that $\mu$ is not freely available
	to be determined by the measured value of $m_Z=91.2$ GeV.
	Thus, the equations \ref{eq:Bmu} and \ref{eq:mzs_ewsb} must be used
	to map out the derived values of $m_Z$ in the $\mu$ vs. $\tan\beta$
	plane.
	
\begin{figure}[!htbp]
		\begin{center}
			\includegraphics[height=0.4\textheight]{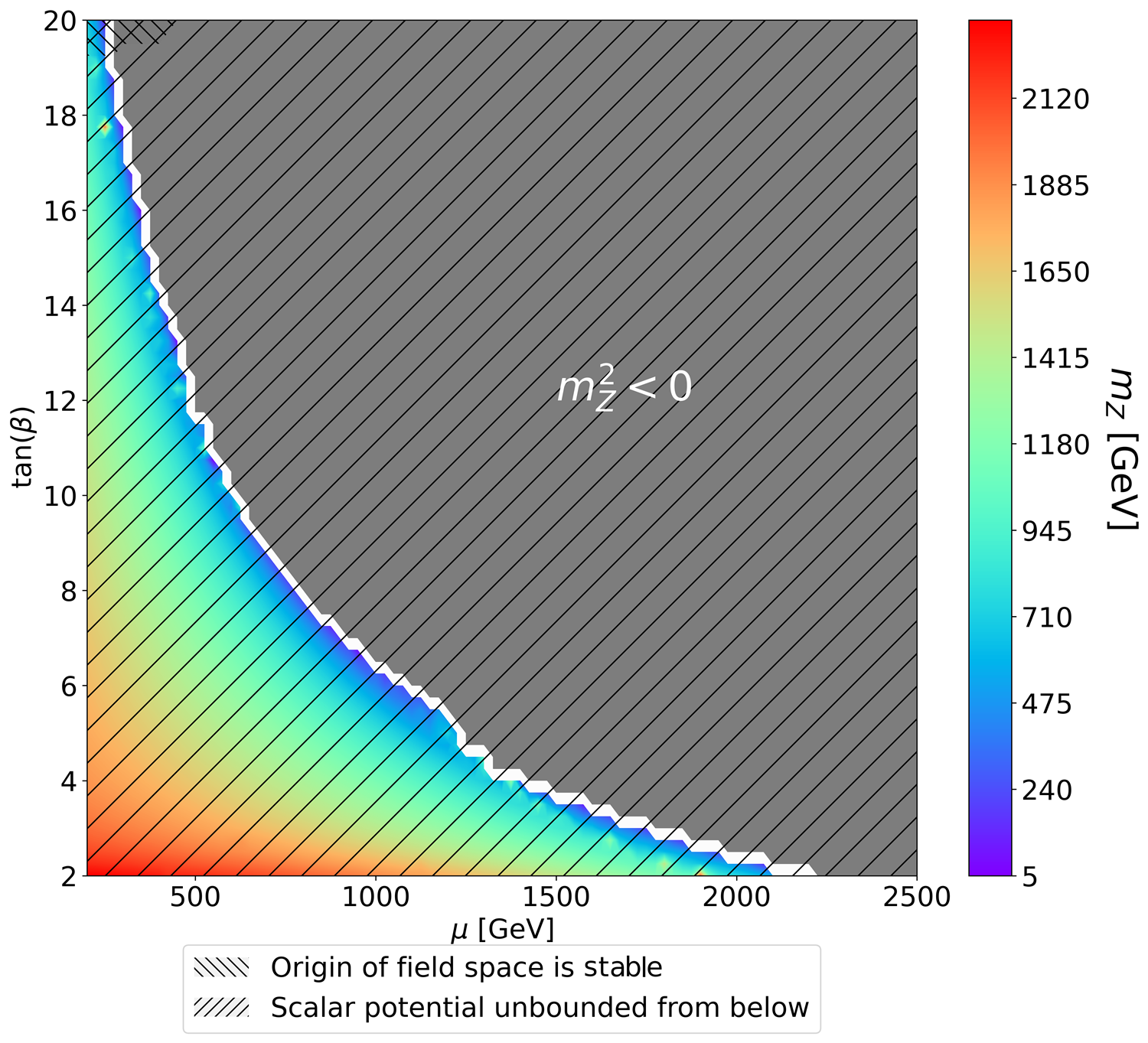}
			\caption{Computed value of $m_Z$ in the $\mu_{GUT}$ vs. $\tan\beta$ plane
				for the PRS BM point with $m_{1/2}=-A_0=1.5$ TeV, and $M_{\text{int}}=10^{11}$ GeV.
				\label{fig:mZ11}}
		\end{center}
	\end{figure}

	This is shown in Fig. \ref{fig:mZ11} for the case at hand.
	Here, we see that for large $\mu$ values, then $m_Z^2$ is computed at loop level to be negative.
	For smaller $\mu$, then typically $m_Z$ is of order the TeV scale.
	For a given value of $\tan\beta$, one can choose $\mu$ near the edge
	of the gray excluded region where $m_Z\sim 100$ GeV.
	For Fig. \ref{fig:RGE11}, we have chosen
	$\tan\beta =10$ which then fixes $\mu\sim 500$ GeV. Unfortunately, for
	all choices of $\mu$ and $\tan\beta$ shown in the plane, we find the scalar
	potential to be UFB in the $D$-flat direction.

	\subsection{Results for $A_0>0$}
	
\begin{figure}[!htbp]
		\begin{center}
			\includegraphics[height=0.4\textheight]{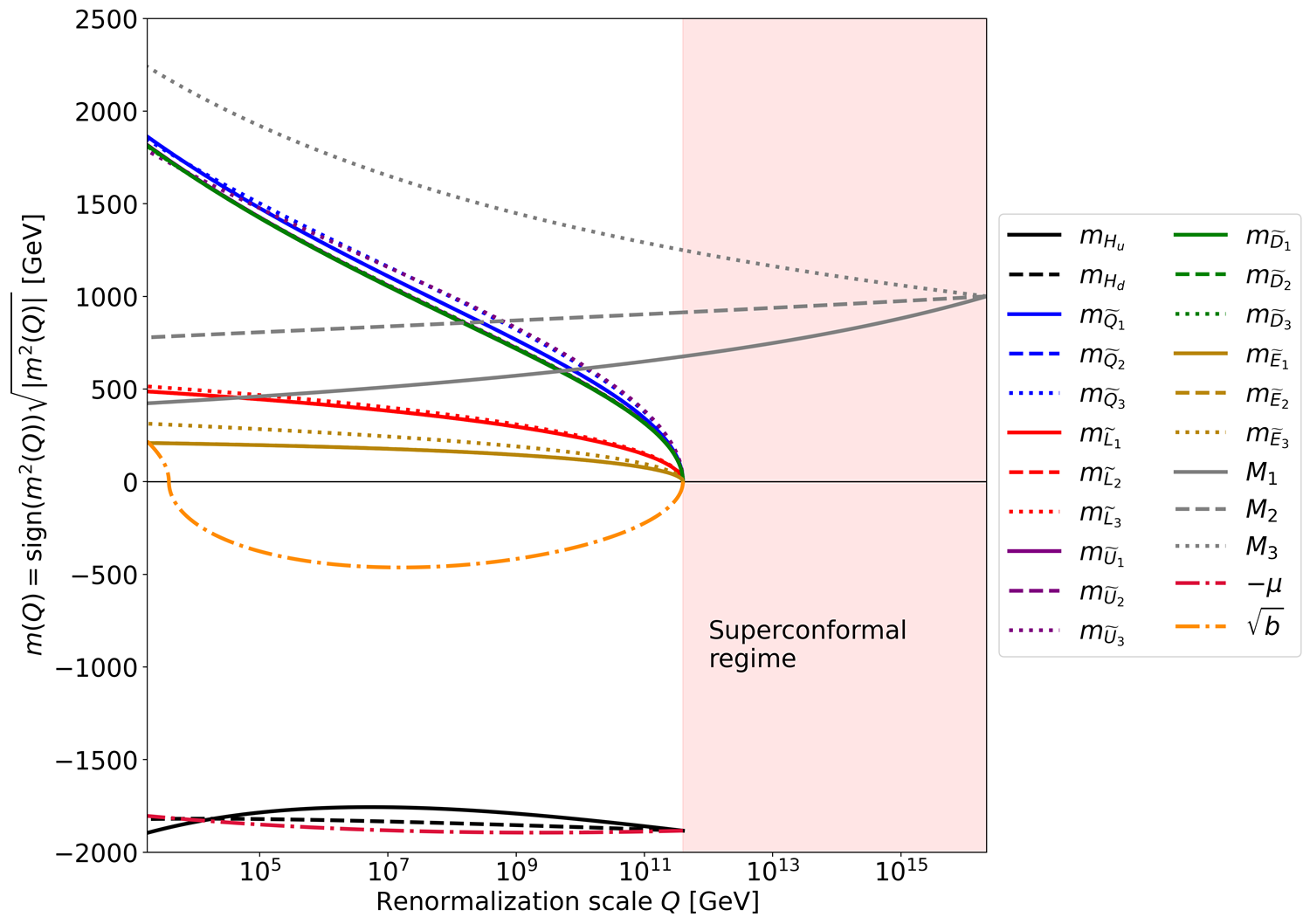}\\
			\includegraphics[height=0.4\textheight]{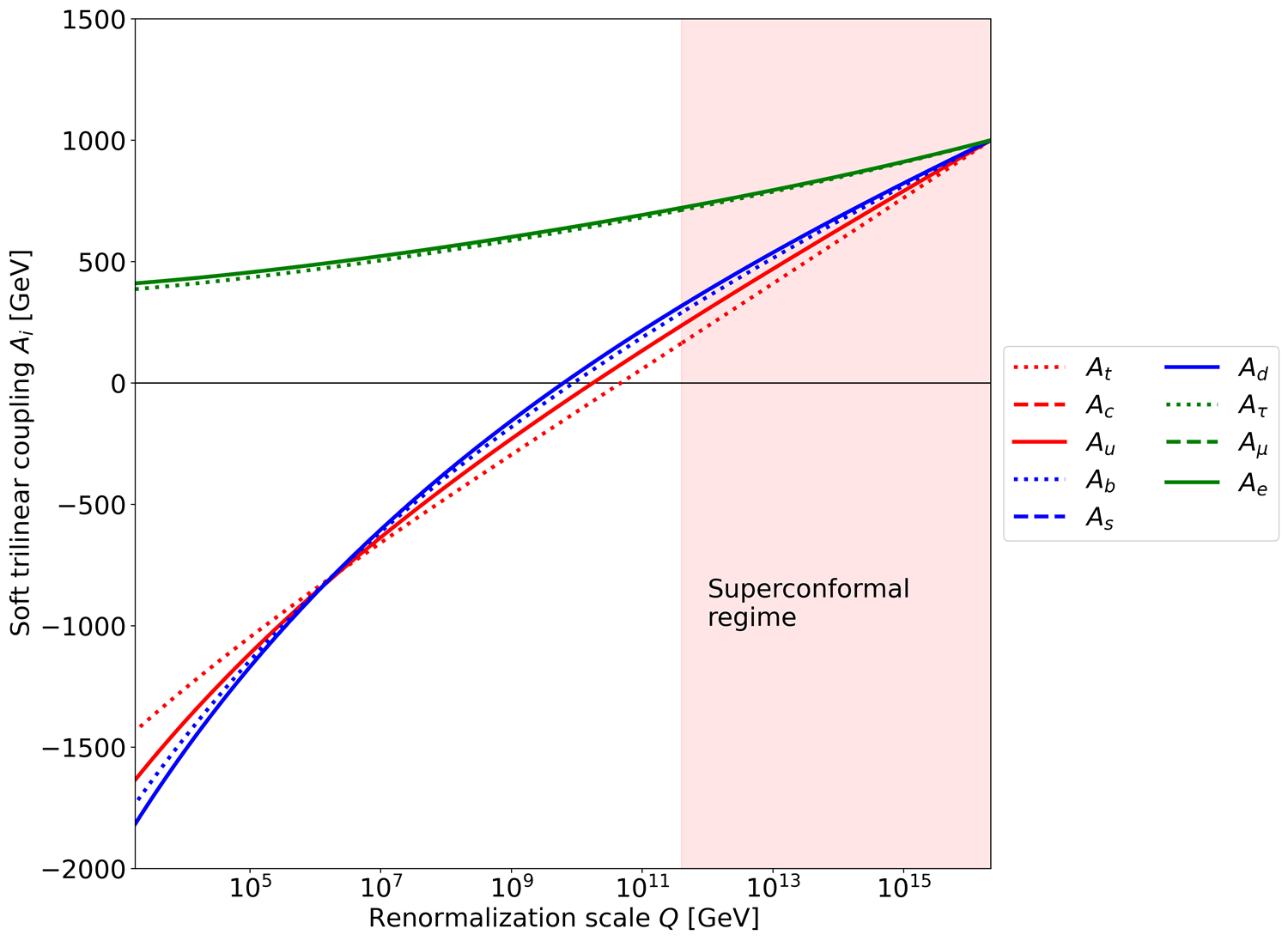}
			\caption{Running of soft terms and $\mu$ in the PRS
				scalar sequestering scheme for $m_{1/2}=1$ TeV, $A_0=m_{1/2}$,
				and $\mu = 1.8$ TeV. We take the intermediate scale
				$M_{int} =4\times 10^{11}$ GeV.
				In frame {\it a}) we show running scalar masses
				and $\mu$ term while in frame {\it b}) we show the running trilinear
				soft terms.
				\label{fig:RGEAgt0}}
		\end{center}
	\end{figure}

\begin{figure}[!htbp]
		\begin{center}
			\includegraphics[height=0.35\textheight]{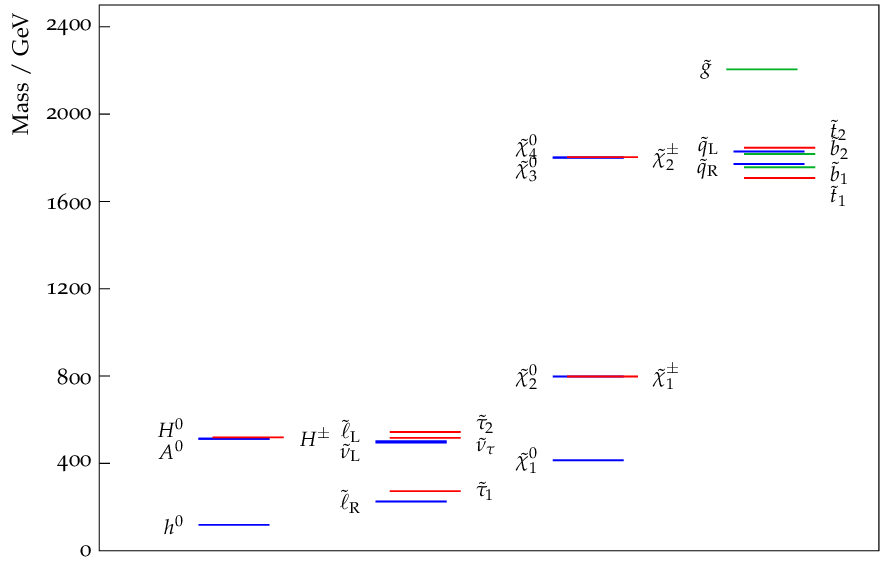}
			\caption{The resulting mass spectrum with a characteristic slepton as the LSP for the PRS scheme with $m_{1/2}=1$ TeV, $A_0=m_{1/2}$,
				and $\mu = 1.8$ TeV. The spectrum was produced using SoftSUSY v4.1.17 \cite{Allanach:2002ss} and slhaplot \cite{Buckley:2015py}.}
			\label{fig:PRS_slepLSP_spec}
		\end{center}
	\end{figure}

	In Fig. \ref{fig:RGEAgt0}, we show a PRS point that does develop appropriate
	EWSB where $M_{int} =4\times 10^{11}$ GeV and $m_{1/2}=A_0=1$ TeV.
	For $\tan\beta =21.25$, we find $\mu \simeq 1.8$ TeV.
	In this case, with $A_0=1$ TeV, we see from frame {\it b}) that the $A_i$ parameters
	are all positive for large $Q$, with $A_t$ and $A_b$ becoming small and then
	negative around $Q\alt  10^{10}$ GeV. This feeds into the $b$ parameter
	evolution causing $b$ to run at $Q< M_{int}$ to negative values until the
	large negative $A_i$ terms cause it to turn up and become positive around $Q\sim m_{weak}$,
	aiding in appropriate EWSB.
	While this model does develop a viable EW vacuum, the slepton masses evolve only to
	$m_{E_i}\sim 250$ GeV at $Q\sim m_{weak}$ so that slepton masses are well below
	both the $\mu$ and $M_1$ terms. Thus, for this point we have a charged slepton as the
	lightest SUSY particle.
	The derived sparticle mass spectra for this case are shown in Fig. \ref{fig:PRS_slepLSP_spec}.

	In the case shown, with the MSSM-only as the low energy EFT, then
	one would expect a charged stable relic, and dark matter wouldn't be dark.
	One can circumvent this situation by adding extra particles or interactions
	to the low energy EFT.
	An example of the former would be to add a Peccei-Quinn (PQ) sector with an
	axino $\ta$ as the LSP so that $\tilde{e}_R\to e\ta$. In this case, one would get a
	potentially long-lived but unstable slepton and one must avoid collider
	and other constraints on such objects. The slepton lifetime would depend on the
	assumed value of the PQ scale $f_a$. An example of added interactions would be
	to postulate broken $R$-parity so that the slepton LSP decays to SM particles.
	Then one must explain why some RPV couplings are substantial whilst others
	are very small, as required by proton stability bounds\cite{Barger:1989rk,Dreiner:1997uz}.
	
	\section{Parameter space scans: PRS scheme}
	\label{sec:pspace}

        \subsection{Universal gaugino masses}
        
	In order to search for viable weak scale SUSY spectra in the PRS scheme, we
	implement a scan over the PRS parameter space:
	\bi
	\item $m_{1/2}:\ 0.2\rightarrow 5$ TeV
	\item $A_0:\ -5 \rightarrow +5$ TeV
	\item $M_{int}:\ 10^6\rightarrow 10^{14}$ GeV.
	\ei
	Our code then scans over values of $(\mu ,\tan\beta )$ leading to $m_Z\sim 91.2$ GeV.
	We then check for CCB minima, points that are UFB, and appropriate EWSB.
	For points that pass all criteria with appropriate EWSB, we then check for
	a neutral or a charged LSP.
	
\begin{figure}[!htbp]
		\begin{center}
			\includegraphics[height=0.4\textheight]{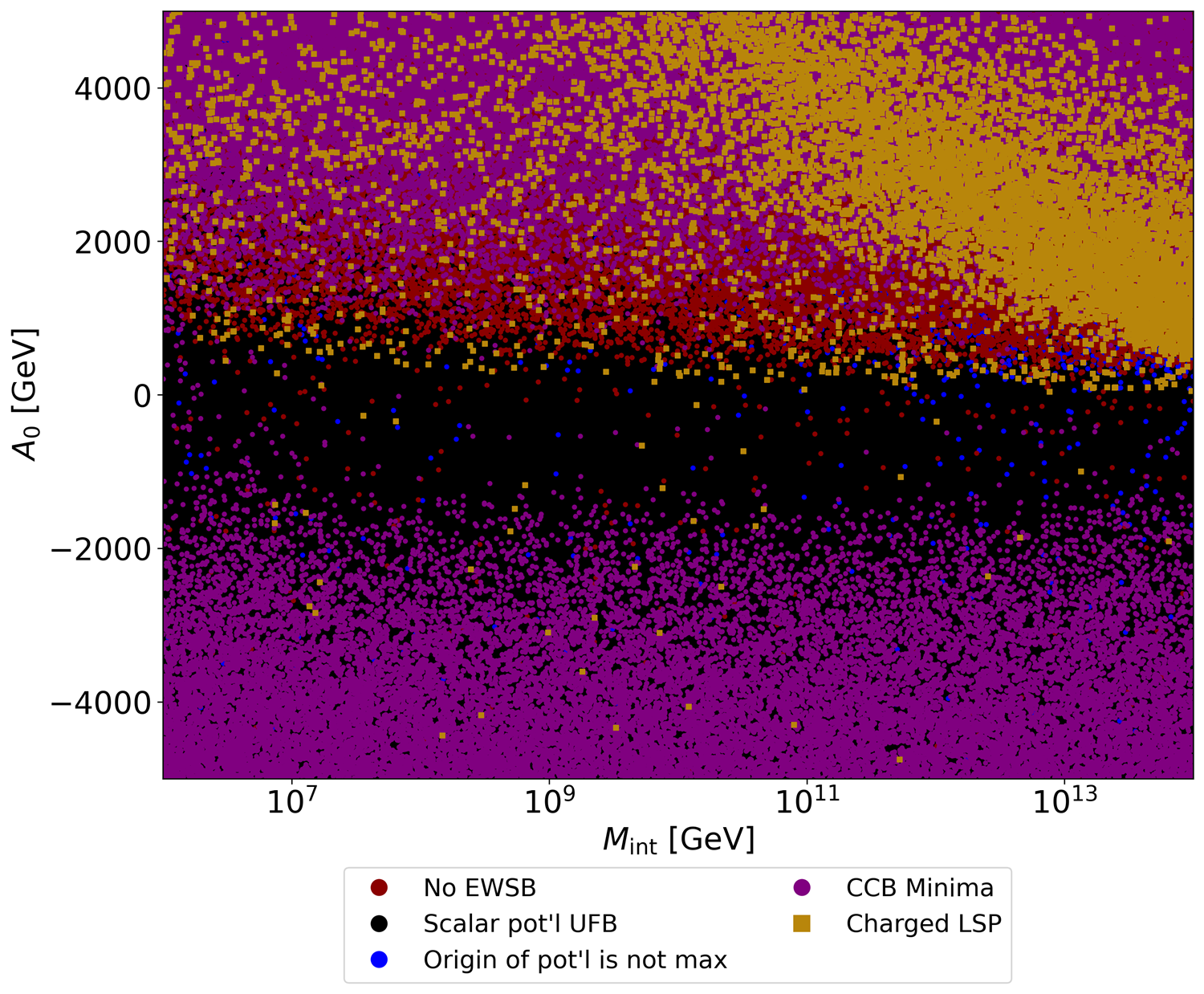}\\
			\includegraphics[height=0.4\textheight]{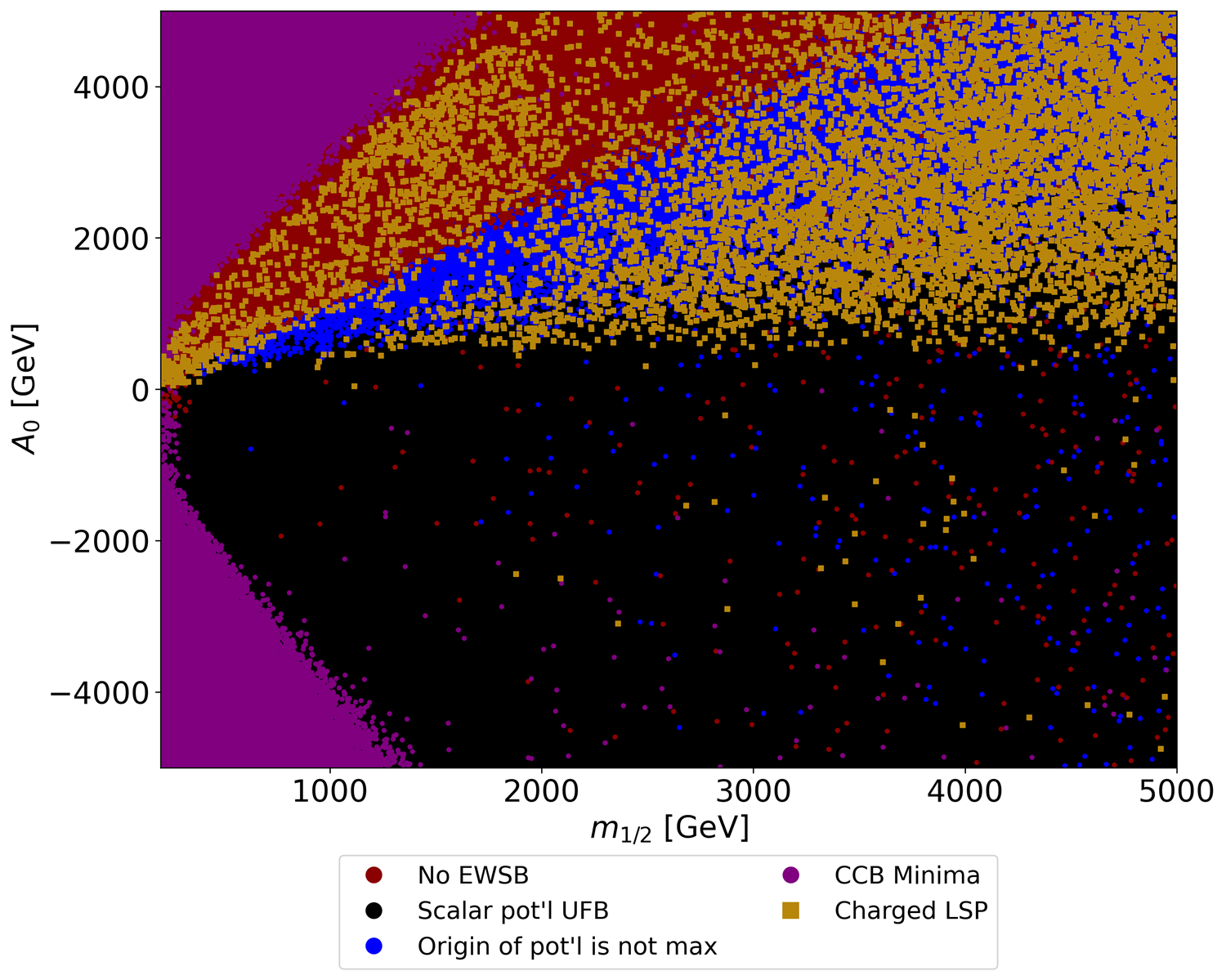}
			\caption{Scan over the PRS parameter space with UGMs in the
				{\it a}) $A_0$ vs. $M_{int}$ plane and {\it b}) the
				$A_0$ vs. $m_{1/2}$ plane.
				\label{fig:PRSscatter1}}
		\end{center}
	\end{figure}

	Our first results are shown in Fig. \ref{fig:PRSscatter1} where we
	show scan points {\it a}) in the $A_0$ vs. $M_{int}$ plane and {\it b})
	in the $A_0$ vs. $m_{1/2}$ plane. From frame {\it a}), we see that only the yellow points
	satisfy all EWSB constraints, although all the surviving points have a
	slepton as the LSP. In particular, the $A_0<0$ points almost all have either CCB minima
	(for large negative $A_0$) or else an UFB potential. For $A_0>0$, then the
	scalar potential is better behaved but frequently does not have appropriate EWSB.
	The scan points with appropriate EWSB are much more prominent at large $M_{int}$ and
	large $m_{1/2}$.

\begin{figure}[!htbp]
		\begin{center}
			\includegraphics[height=0.4\textheight]{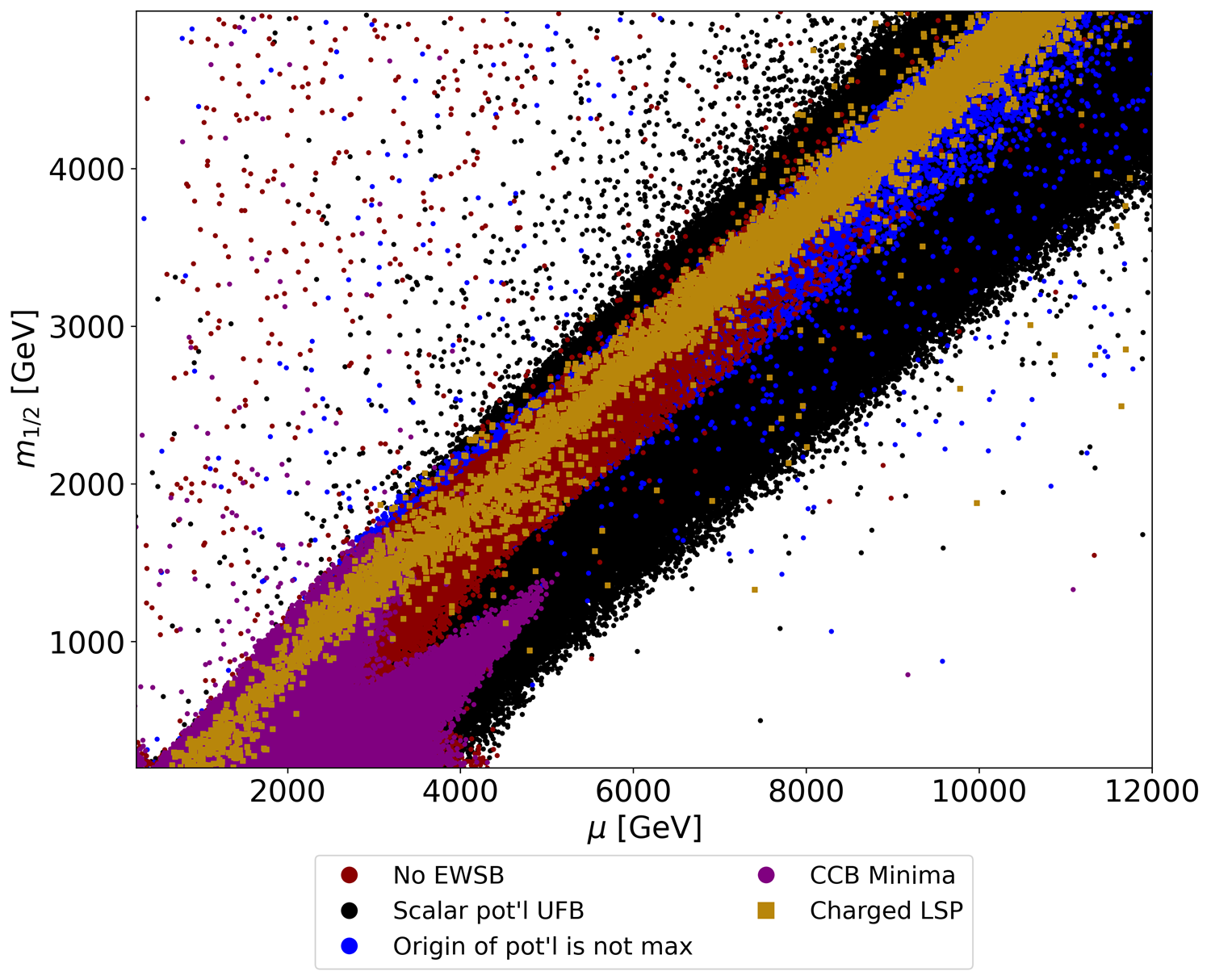}
			\caption{Scan over the PRS parameter space with UGMs in the
				$m_{1/2}$ vs. $\mu$ plane.
				\label{fig:PRSscatter2}}
		\end{center}
	\end{figure}

	In Fig. \ref{fig:PRSscatter2},
	we show our scan points in the  $m_{1/2}$ vs. $\mu$ plane.
	Here, we see some structure where $\mu\sim 2m_{1/2}$ is favored.
	These qualitative features were also found by Perez, {\it et al.} in
	Ref. \cite{Perez:2008ng} where most of their parameter space was excluded
	by EWSB constraints except for large $M_{int}$ where they also found
	$\mu\sim 2m_{1/2}$ and for their lone sample point, they also obtained a
	slepton as the LSP.

	Given our overall scan results in the PRS scheme, we find the strong scalar
	sequestering scenario (with unified gaugino masses) rather difficult
	(but not impossible) to accept:
	the bulk of p-space points have problematic EWSB and any surviving points have a
	charged LSP thus requiring new particles and/or new interactions to evade
	cosmological constraints on charged relics from the Big Bang.
	
	\subsection{Non-universal gaugino masses}

        \subsubsection{Varying $M_1$ and $M_2$}

\begin{figure}[!htbp]
		\begin{center}
			\includegraphics[height=0.4\textheight]{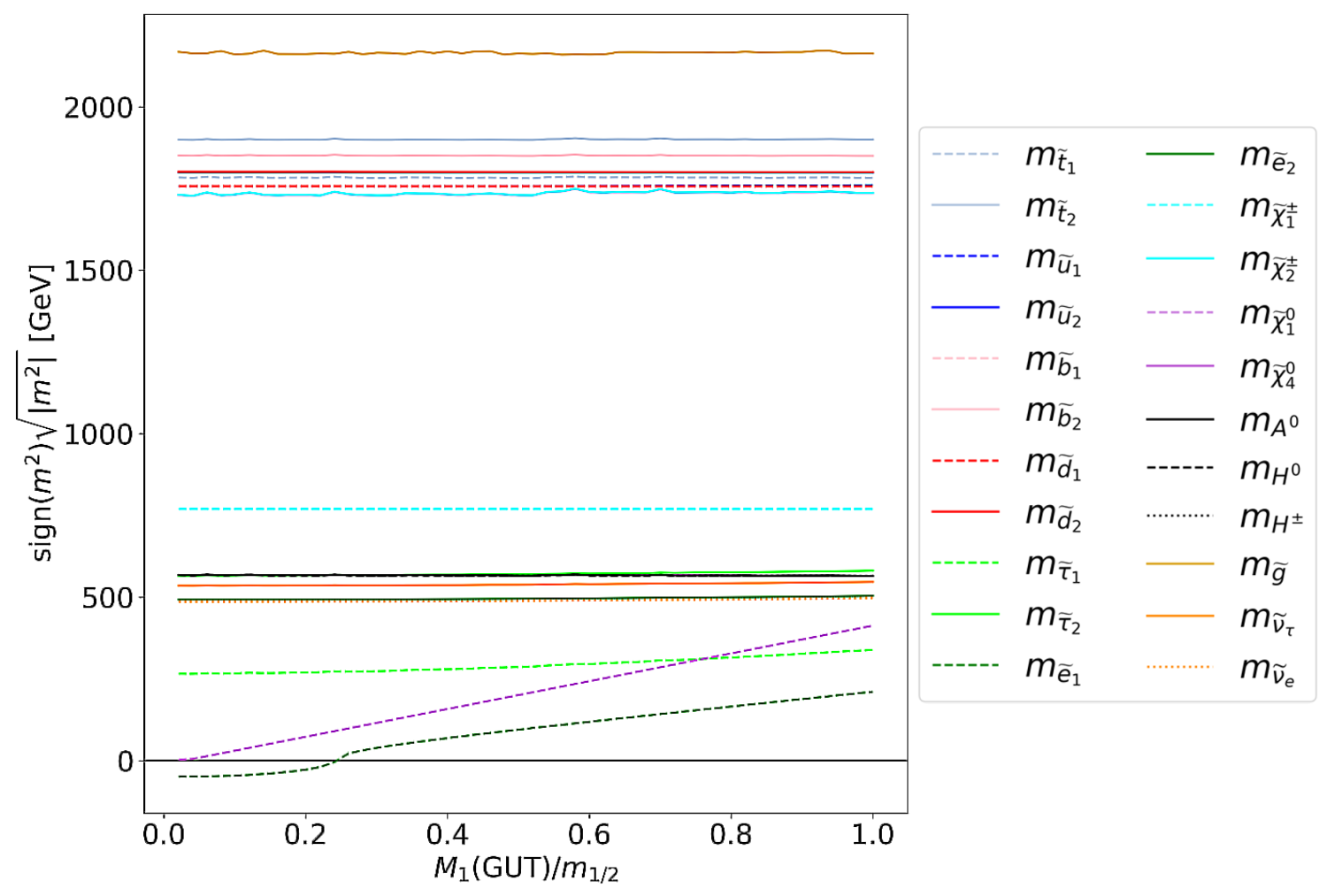}\\
			\includegraphics[height=0.4\textheight]{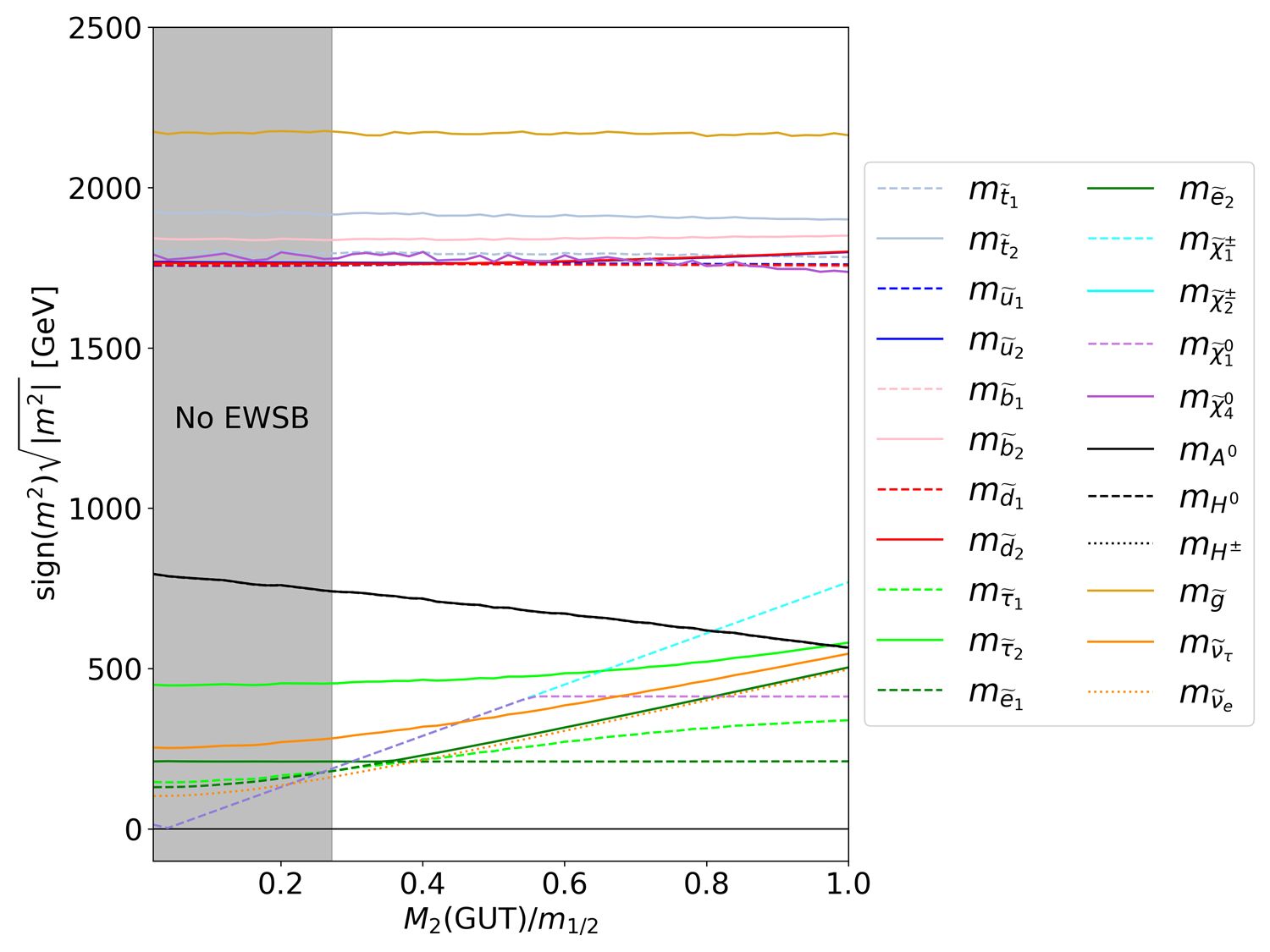}
			\caption{The SUSY mass spectrum vs. GUT-scale gaugino mass
                          parameters $M_{1},M_{2}$ in the PRS model varied below $m_{1/2}$.
                          In both frames, the spectrum at the far right is similar to the spectrum seen in Fig. \ref{fig:PRS_slepLSP_spec}. In frame {\it a}) the mass spectrum as $M_{1}(\text{GUT})$ is varied below $m_{1/2}$ to zero is plotted.
                          The neutralino never becomes the LSP, as the selectron and smuon remain
                          lighter until CCB minima are realized.
                          In frame {\it b}) we display the mass spectrum as $M_{2}(\text{GUT})$ is
                          varied below  $m_{1/2}$ to zero.
                          Near $M_{2}\sim0.4 m_{1/2}$, the sneutrino briefly becomes the LSP before the
                          Higgs potential becomes unbounded from below due to a lack of running
                          in the $b=B\mu$ parameter.
                          Thus, a neutralino LSP cannot be achieved here.
                          In both frames, we take $m_{1/2}=M_{3}(\text{GUT})=A_{0}=1$ TeV,
                          $M_{\text{int}}=4\cdot 10^{11}$ GeV, and $\tan(\beta)=21.25$.}
			\label{fig:PRS_NUGM_scan_spec}
		\end{center}
	\end{figure}

\begin{figure}[!htbp]
		\begin{center}
			\includegraphics[height=0.5\textheight]{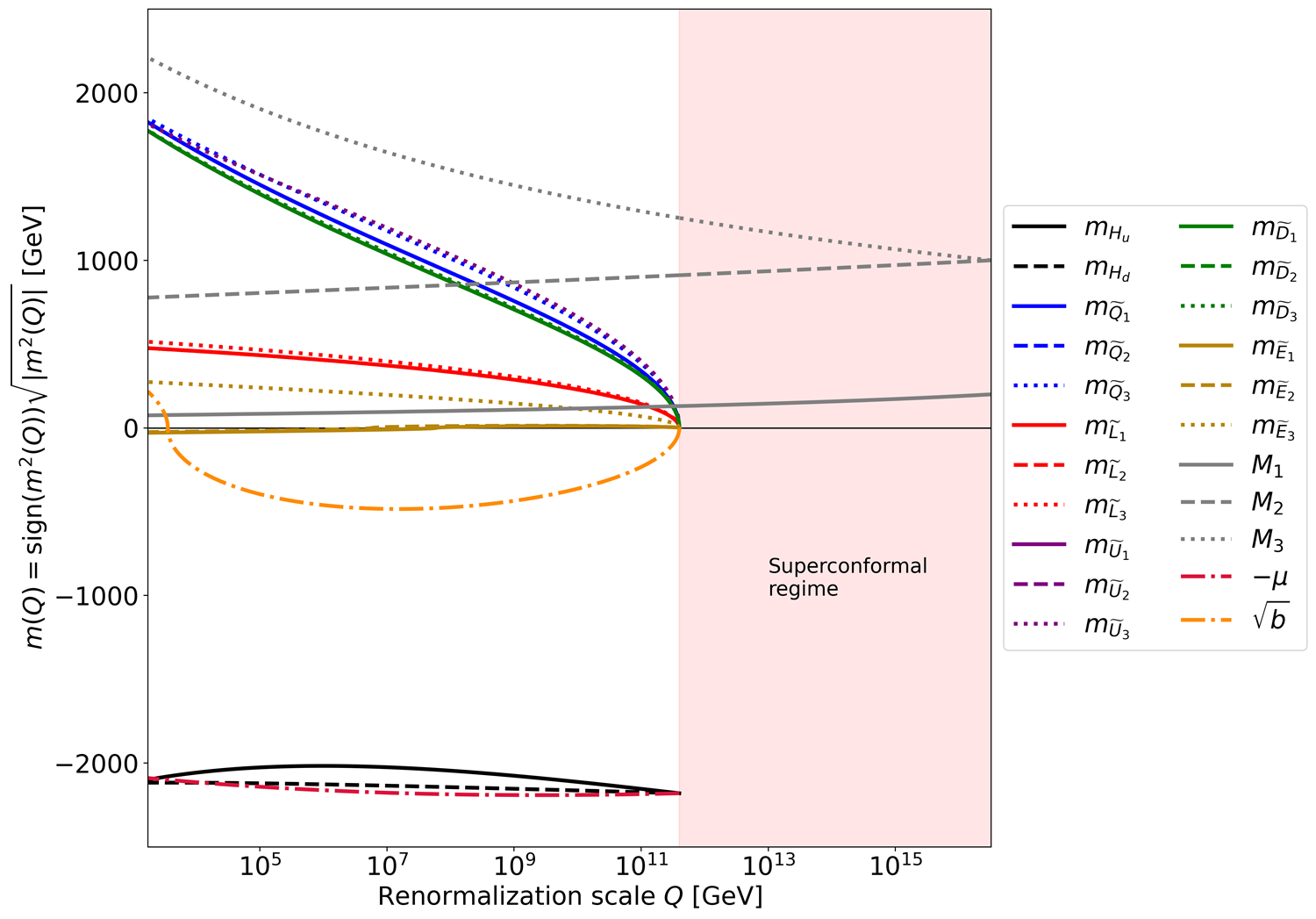}
			\caption{Example RGE running of the soft masses from
                          Fig. \ref{fig:PRS_NUGM_scan_spec}{\it a}) demonstrating the CCB
                          nature of a point with $m_{\widetilde{e}_{R}}^{2}<0$ with
                          $M_{1}(\text{GUT})\sim0.15 m_{1/2}$.
                          Though the left-handed slepton states (red) evolve to moderate values,
                          the right-handed slepton states of the first two generations evolve to be
                          negative at the SUSY scale due to the small value of $M_{1}$.}
			\label{fig:PRS_NUGM_scan_spec_lowM1}
		\end{center}
	\end{figure}

        One possibility to try to circumvent the slepton-LSP problem in the PRS scheme is to appeal
        to NUGMs, by dialing down either $M_1$ or $M_2$ from their unifed values
        until either the bino or the wino becomes the LSP. The computed sparticle mass spectra are shown
        in Fig. \ref{fig:PRS_NUGM_scan_spec} in frame {\it a}) for varying $M_1$ and in frame {\it b})
        for varying $M_2$. From frame {\it a}), we see that as $M_1$ diminishes, the
        lightest neutralino mass $m_{\tchi_1^0}$ does indeed decrease (moving from unified gaugino
        masses on the right to small $M_1$ on the left as shown by the lavender dashed curve). However, as
        $M_1$ decreases, then upward RGE pull on $m_{Ei}$
        (right-slepton soft mass of generation $i$) from the $U(1)_Y$ gaugino also diminishes,
        and ultimately $m^2_{E_{1,2}}$ go tachyonic around $M_1\sim 0.23 m_{1/2}$.
        Note in this case that the stau soft mass remains larger due to a large negative $X_\tau$
        term in the $m_{E_3}^2$ RGE owing to large negative $m_{H_d}^2$.
        This is shown in Fig. \ref{fig:PRS_NUGM_scan_spec_lowM1} which shows the soft mass running for
        a case with small $M_1$ compared to $m_{1/2}$. This behavior where the bino fails to become
        LSP in the PRS scheme with small $M_1$ appears rather general when we scan over all $M_1$
        values (to be shown shortly).

        Likewise, in frame {\it b}), we take $M_2$ to be its unified value on the right-side of
        the plot, and then dial its value down to try to gain a wino as LSP.
        Around $M_2\sim 0.58 m_{1/2}$, the $m_{\tw_1}$ and $m_{\tz_1}$ mass curves coincide, showing that
        the lightest neutralino has gone from bino to wino. However, in this case, the
        right-sleptons remain LSP until $M_2\sim 0.35 m_{1/2}$ whence the left sleptons, and particularly here
        the left-sneutrino, becomes LSP. Left-sneutrinos have direct detection cross sections
        for scattering on $Xe$ nuclei of $\sigma (\tnu_{eL} Xe\to \tnu_{eL} Xe)$ of
        $\sim 4.5\times 10^{-23}$ cm$^2$\cite{Srednicki:1986vj}, about 23 orders of magnitude large
        than current LZ limits\cite{LZ:2022lsv}, and so are excluded as dark matter.
        For somewhat lower values of $M_2$,
        then $B\mu$ runs to very small values, leading to a UFB scalar potential. This behavior
        also seems rather general from our PRS scan with NUGMs.

\subsubsection{Scan over PRS scheme with NUGMs}
        
        For completeness in our search for viable weak scale SUSY spectra in the PRS scheme, we
	can adopt the case of non-universal gaugino masses
        and scan over this expanded parameter space:
	\bi
	\item $M_{1}(\text{GUT}):\ 0.2\rightarrow 5$ TeV
	\item $M_{2}(\text{GUT}):\ 0.2\rightarrow 5$ TeV
	\item $M_{3}(\text{GUT}):\ 0.2\rightarrow 5$ TeV
	\item $A_0:\ -5 \rightarrow +5$ TeV
	\item $M_{int}:\ 10^6\rightarrow 10^{14}$ GeV.
	\ei
	Similar to above, our code then finds pairs of $(\mu ,\tan\beta )$
        leading to $m_Z\sim 91.2$ GeV.
	We then check for CCB minima, points that are UFB, and appropriate EWSB.
	For points that pass all criteria with appropriate EWSB, we then check for
	a neutral or a charged LSP along with LHC constraints on the gluino mass and
        lightest stop mass. 
        As discussed above, one may try to dial down the $M_{1}(\text{GUT})$ parameter to obtain a
        neutralino LSP, though this leads to both CCB and EWSB issues in this model.
        The issue of a charged LSP persists as in the UGM case,
        though it is possible to accommodate a sneutrino LSP in some cases,
        when $M_{2}(\text{GUT})<M_{1}(\text{GUT})$.
        However, this scenario is severely ruled out due to direct dark matter detection constraints.
	
\begin{figure}[!htbp]
		\begin{center}
			\includegraphics[height=0.4\textheight]{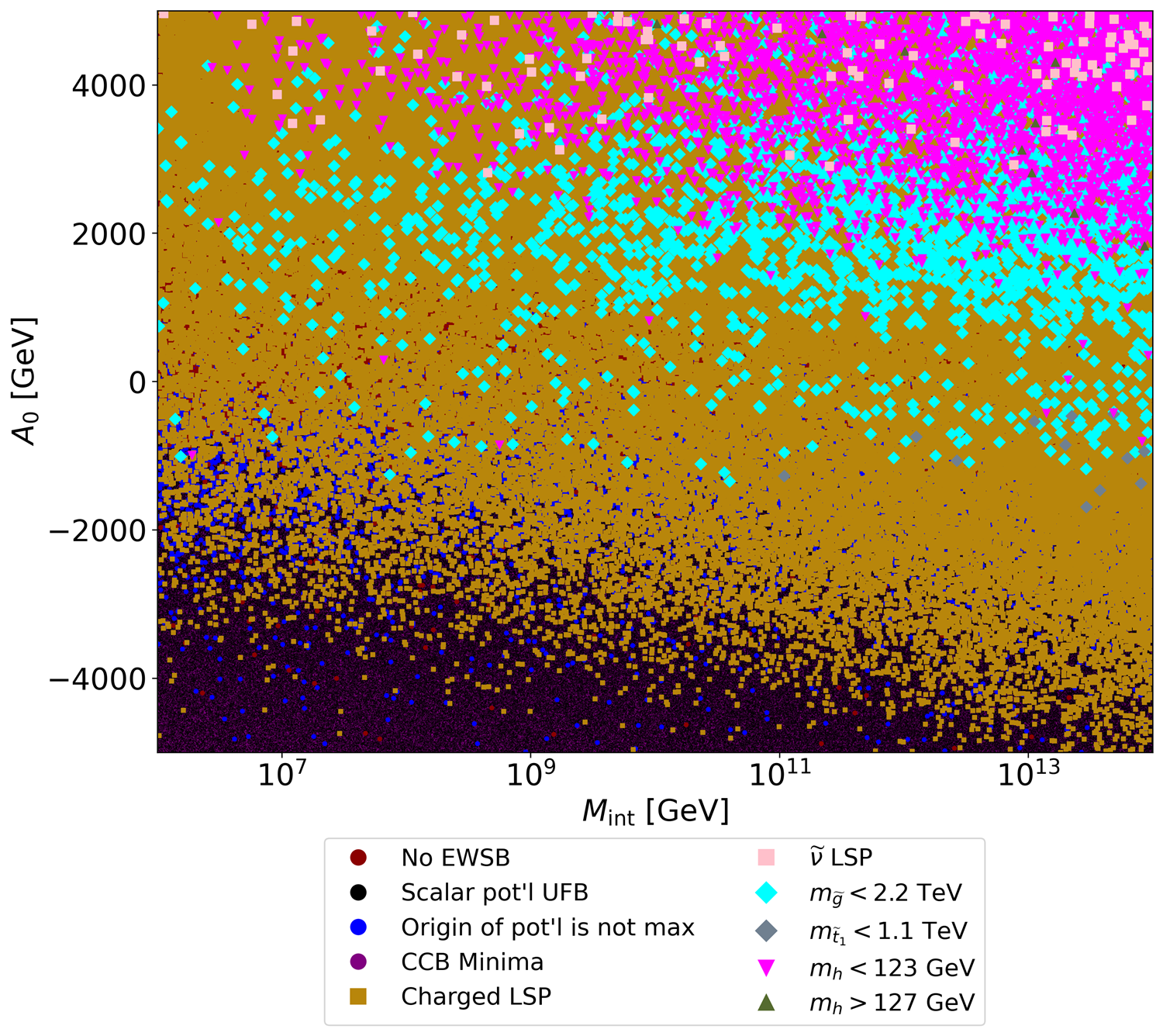}\\
       			\includegraphics[height=0.4\textheight]{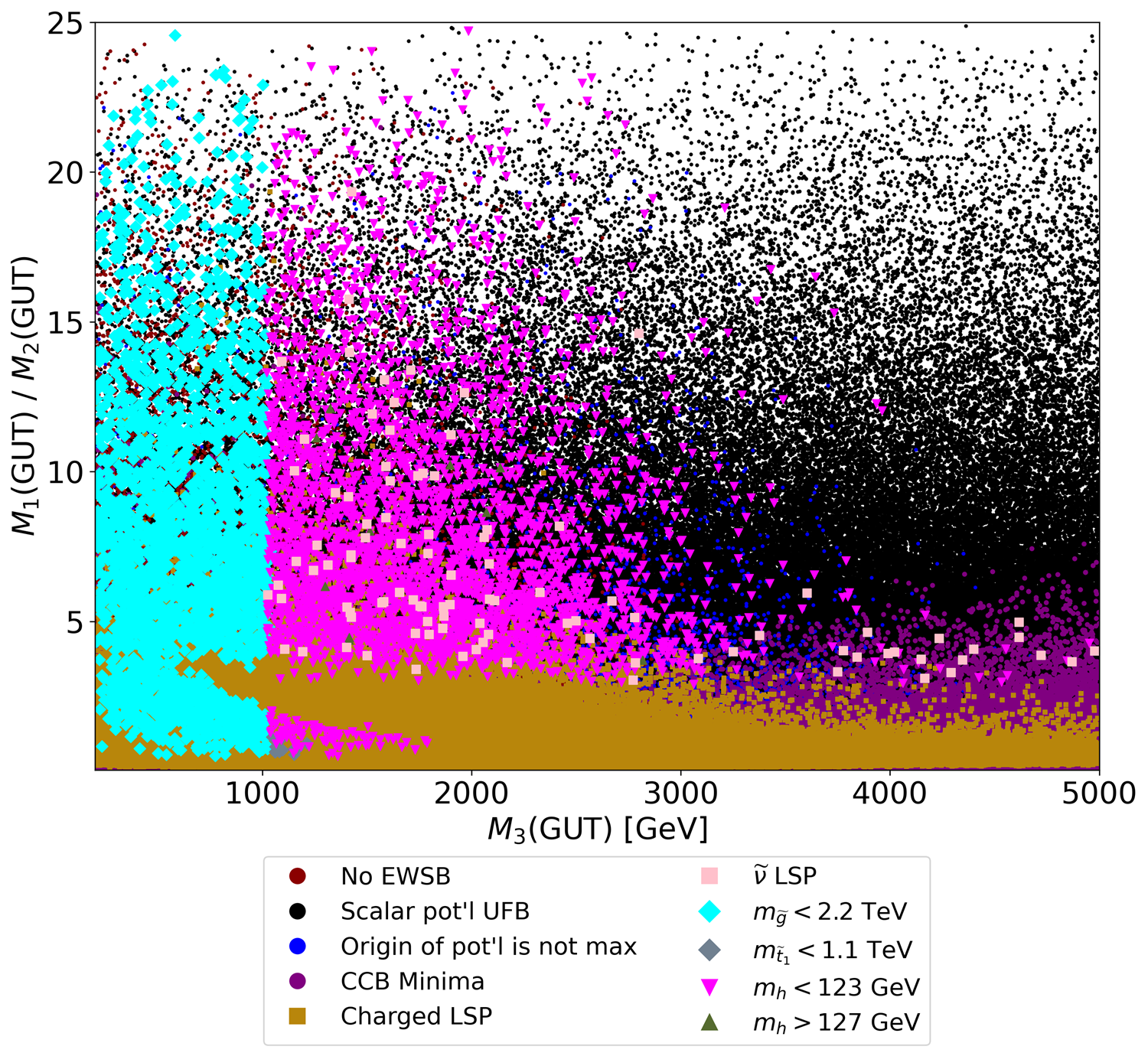}
			\caption{Scan over the PRS parameter space with NUGMs in the
				{\it a}) $A_0$ vs. $M_{int}$ plane and {\it b}) the
				$M_{1}(\text{GUT})/M_{2}(\text{GUT})$ vs. $M_{3}$(GUT) plane.
				\label{fig:PRSNUGMscatter1}}
		\end{center}
	\end{figure}

	Our non-universal gaugino mass scan results are demonstrated in
        Fig. \ref{fig:PRSNUGMscatter1} where we show scan points
        {\it a}) in the $A_0$ vs. $M_{int}$ plane and {\it b}) in the $M_{1}(\text{GUT})/M_{2}(\text{GUT})$ vs. $M_{3}$(GUT) plane. Even with NUGMs, we do not find any points where EWSB is appropriately broken
        but without a charged slepton or left-sneutrino LSP.

	\section{Scalar sequestered SUSY: SPM approach}
	\label{sec:SPM}
	
	In the SPM approach\cite{Martin:2017vlf}, it is noticed that there
	exist bounds on the scaling dimension $\Gamma$ such that $\Gamma$
	is positive but not too large, with $\Gamma\sim 0.3$ maximally\cite{Poland:2011ey,Poland:2015mta,Green:2012nqa}. In this case, the superconformal running
	may be much less, and comparable to the MSSM running.
	Let us denote the dimension 1 soft breaking terms as $m_1$ and
        dimension 2 soft terms as $m_2$.
	Then, after several field rescalings, the
	dimension-1 terms (the $M_i$, $a_{ijk}$ and $\mu$) run according to
	\be
	\frac{dm_1}{dt}=\beta_{m_1}^{MSSM}
	\ee
	while dimension-2 terms (matter scalars $m_{\phi_{ij}}^2$,
	$\hat{m}_{H_{U,d}}$ and $b$) run as
	\be
	\frac{dm_2}{dt}=\Gamma m_2^2+\beta_{m_2}^{MSSM}
	\ee
	where the $\beta^{MSSM}$ are the usual MSSM beta functions and
	$t=\log(Q/Q_0)$ where $Q$ is the energy scale and $Q_0$ is a reference scale.
	For the superconformal regime with $M_{int} <Q<m_*$, then $\Gamma\ne 0$
	but for $Q< M_{int}$, then the superconformal symmetry is broken and
	integrated out, and $\Gamma \to 0$.
	
	An intriguing effect in this case is that the $m_2^2$ terms can run
	until $\Gamma m_2^2\simeq \beta_{m_2}^{MSSM}$ which defines a
	{\it quasifixed point} for the $m_2^2$ running at
	$m_2^2\simeq -\beta_{m_2}^{MSSM}/\Gamma$. Approximate expressions for
	the fixed point values are given by SPM\cite{Martin:2017vlf},
	but will not be repeated here.
	Thus, the $m_2^2$ terms tend to approach their quasifixed point values
	as $Q\to M_{int}$ instead of zero, as in the PRS scheme. This behavior
	helps to ameliorate the problems of the PRS scheme with respect to EWSB.

\begin{figure}[!htbp]
		\begin{center}
		  \includegraphics[height=0.3\textheight]{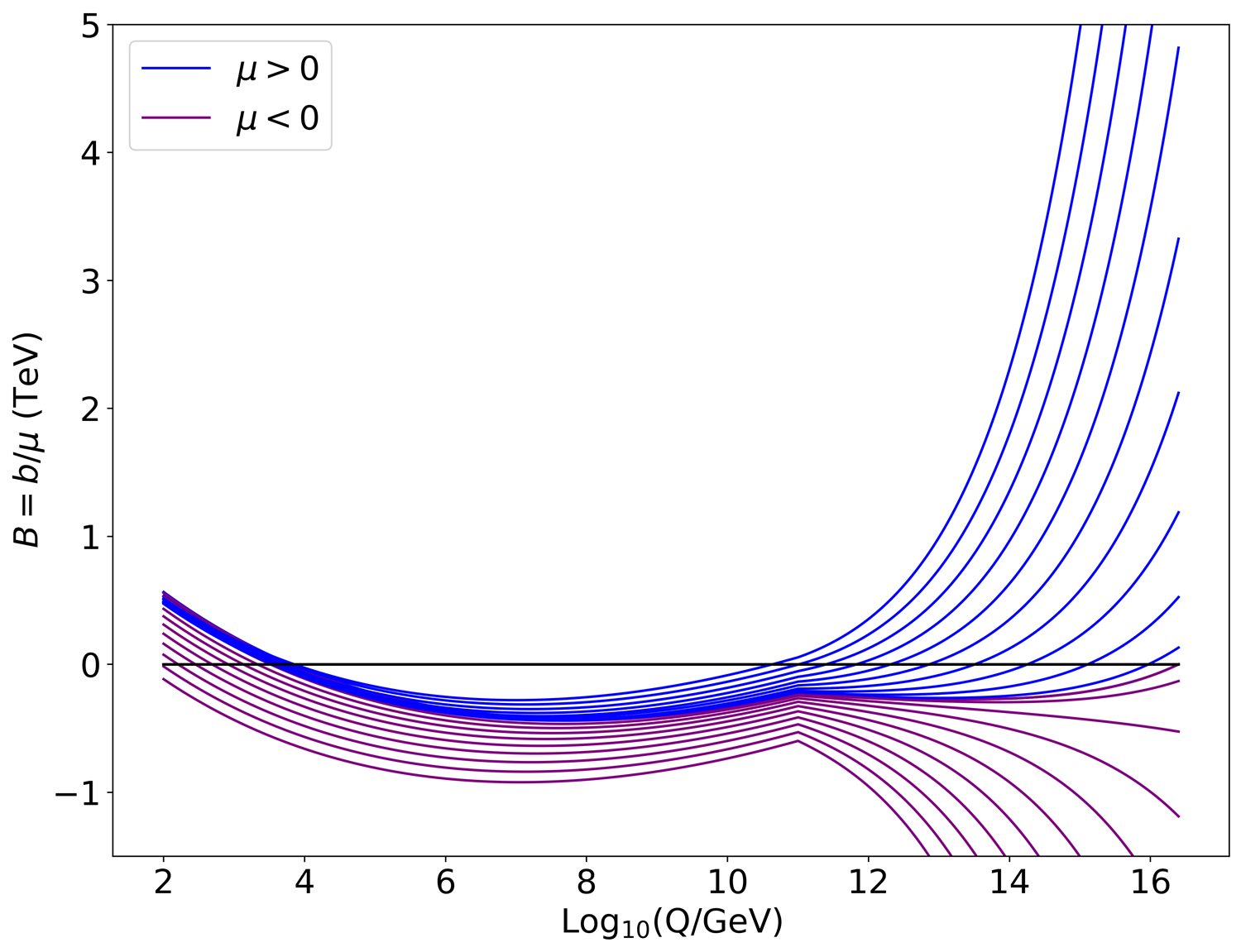}\\
		  \includegraphics[height=0.3\textheight]{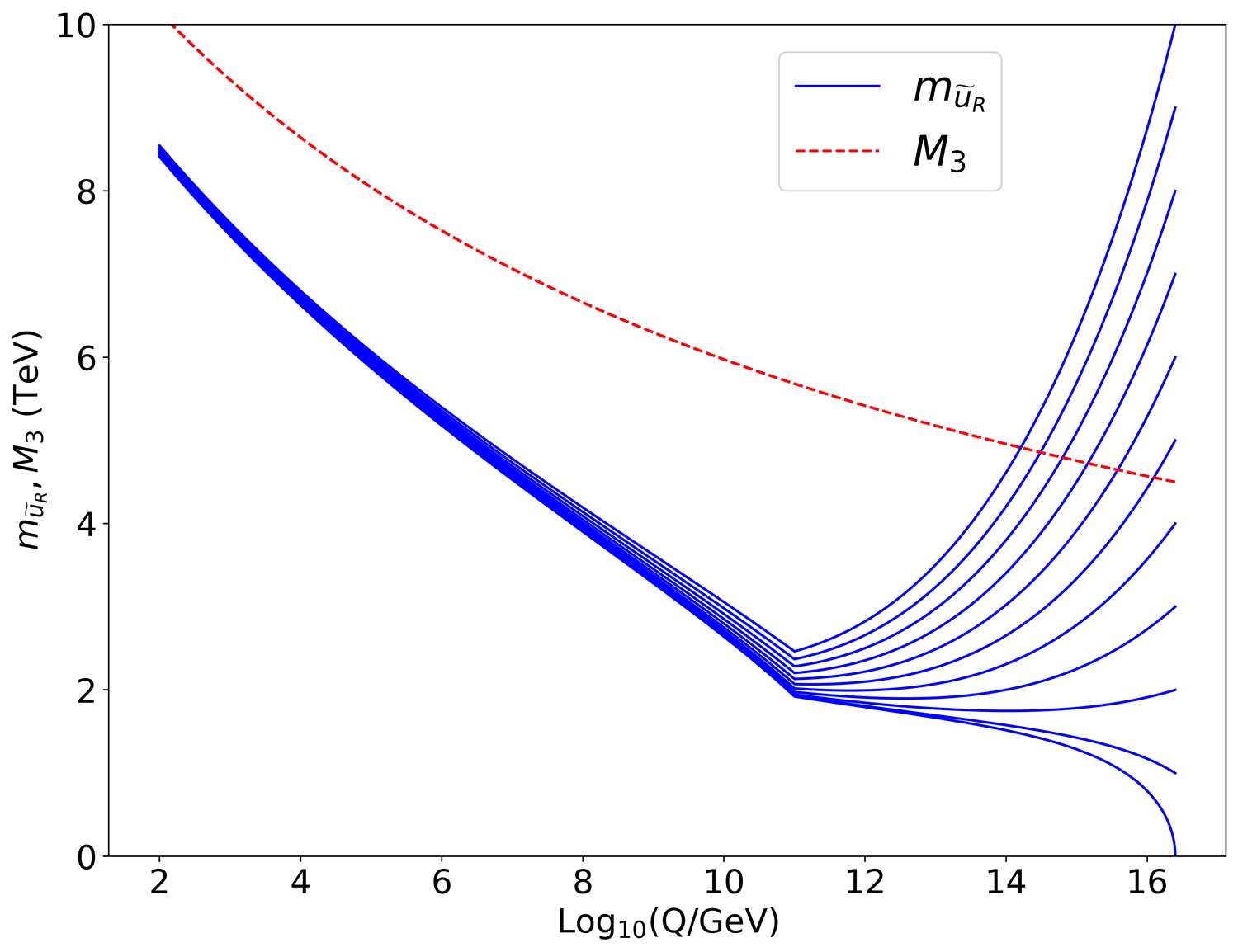}\\
			\includegraphics[height=0.3\textheight]{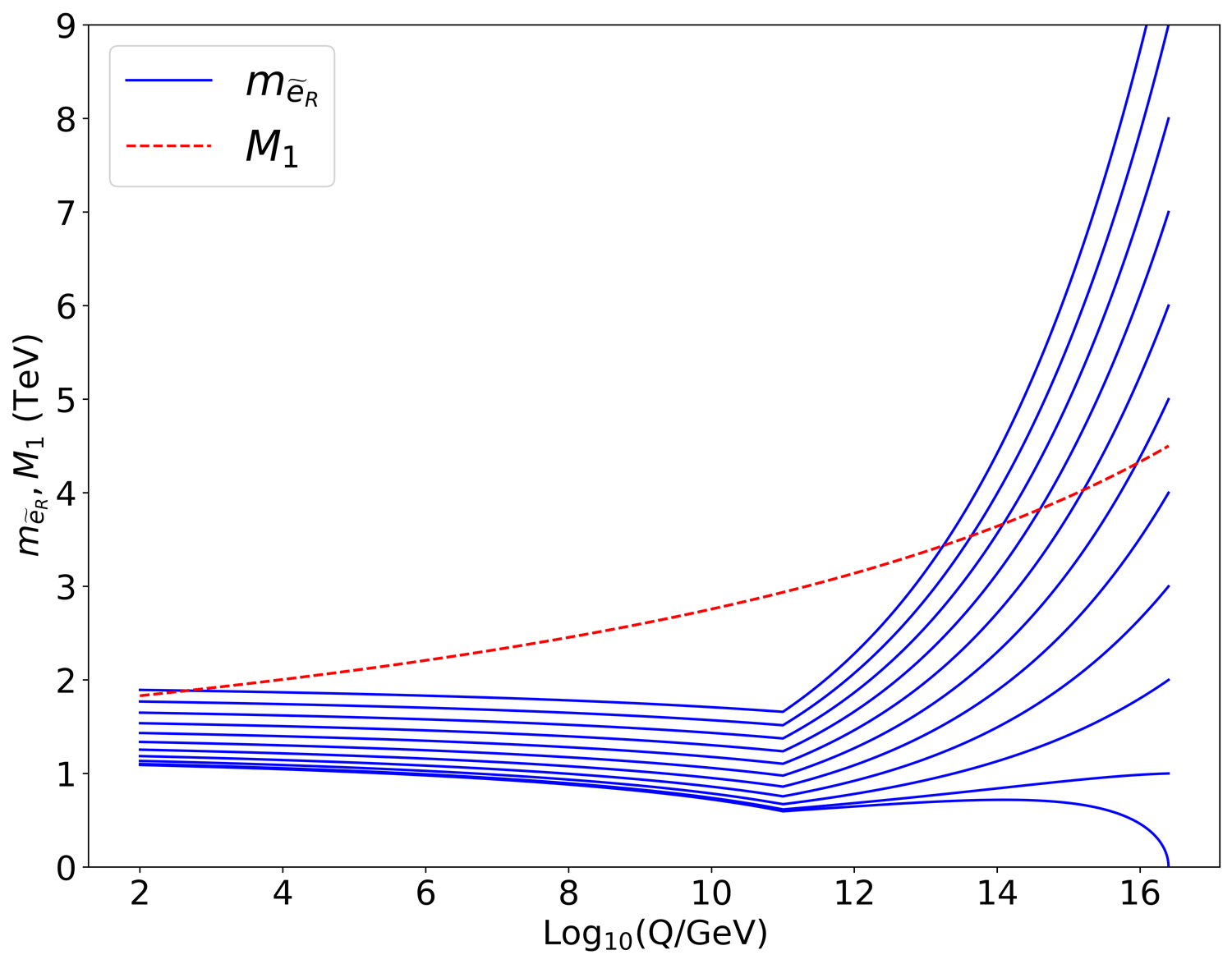}
			\caption{Running of {\it a}) $B\equiv b/\mu$, {\it b}) $m_{U_1}$ and $M_3$
                          and {\it c}) $m_{E_1}$ and $M_1$ from $Q=m_{GUT}$ to
				$Q=m_{weak}$ under the SPM scheme with $\Gamma =0.3$.
				}
		\label{fig:quasi2}
		\end{center}
	\end{figure}

\begin{figure}[!htbp]
		\begin{center}
			\includegraphics[height=0.3\textheight]{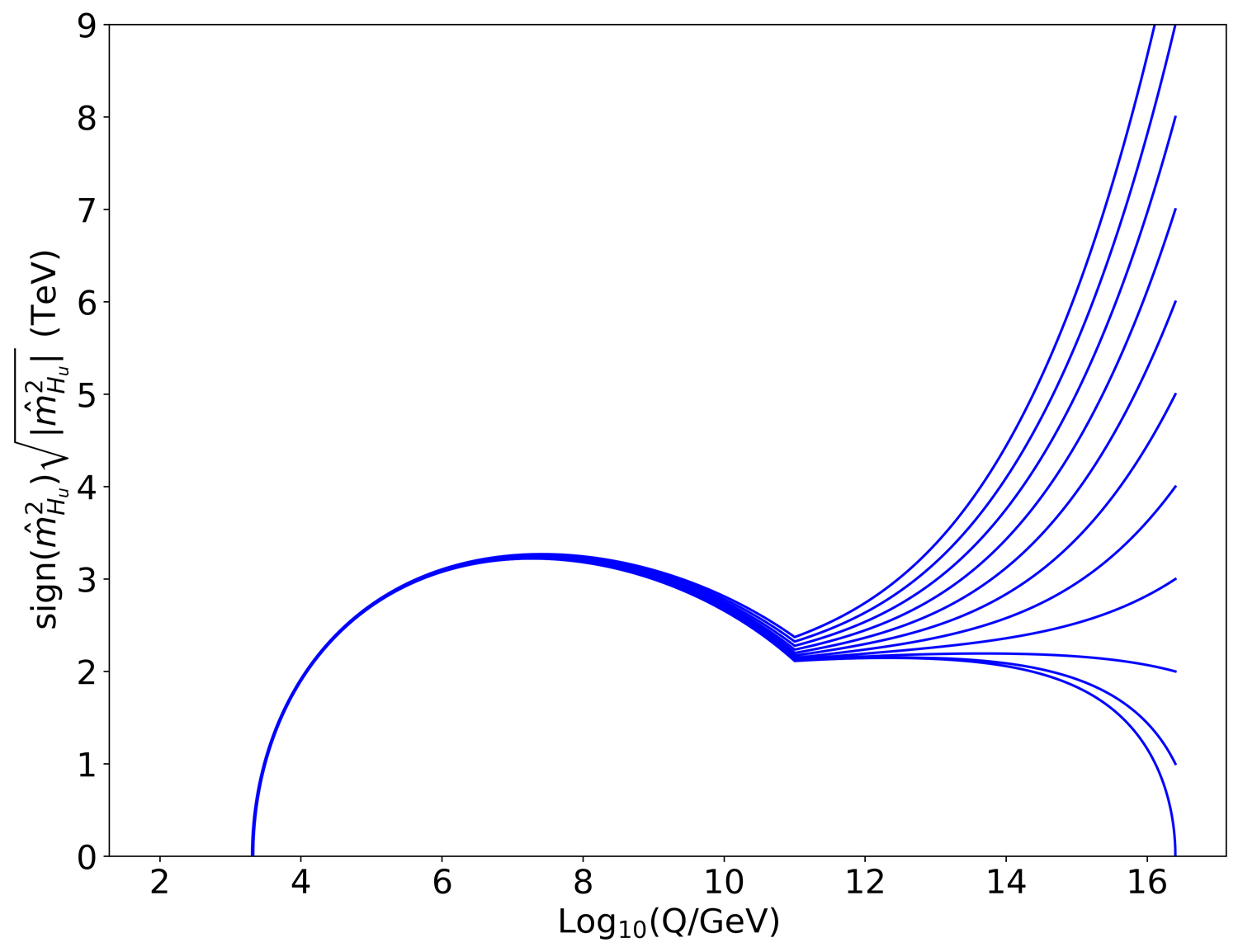}\\
			\includegraphics[height=0.3\textheight]{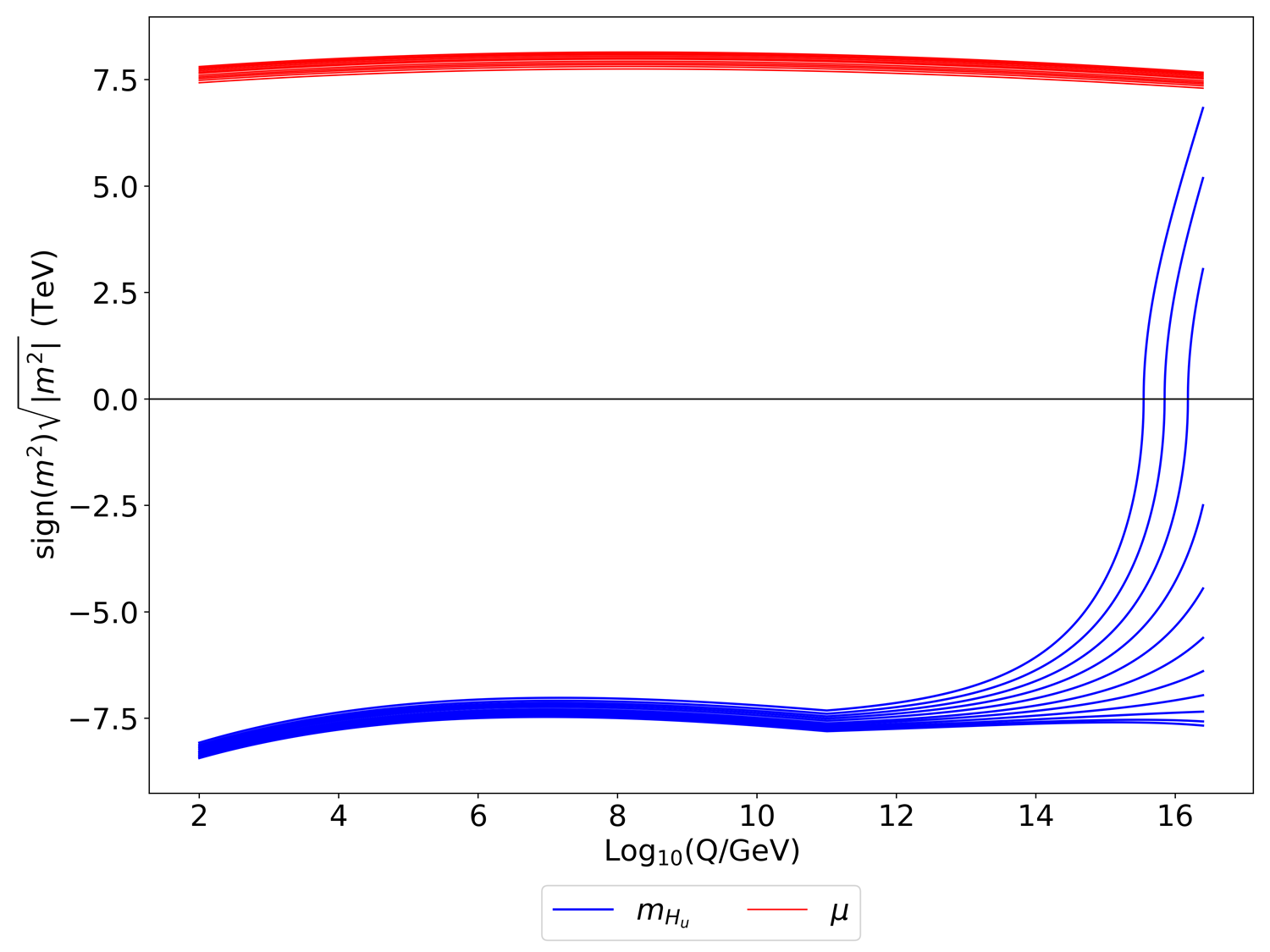}\\
                        \includegraphics[height=0.3\textheight]{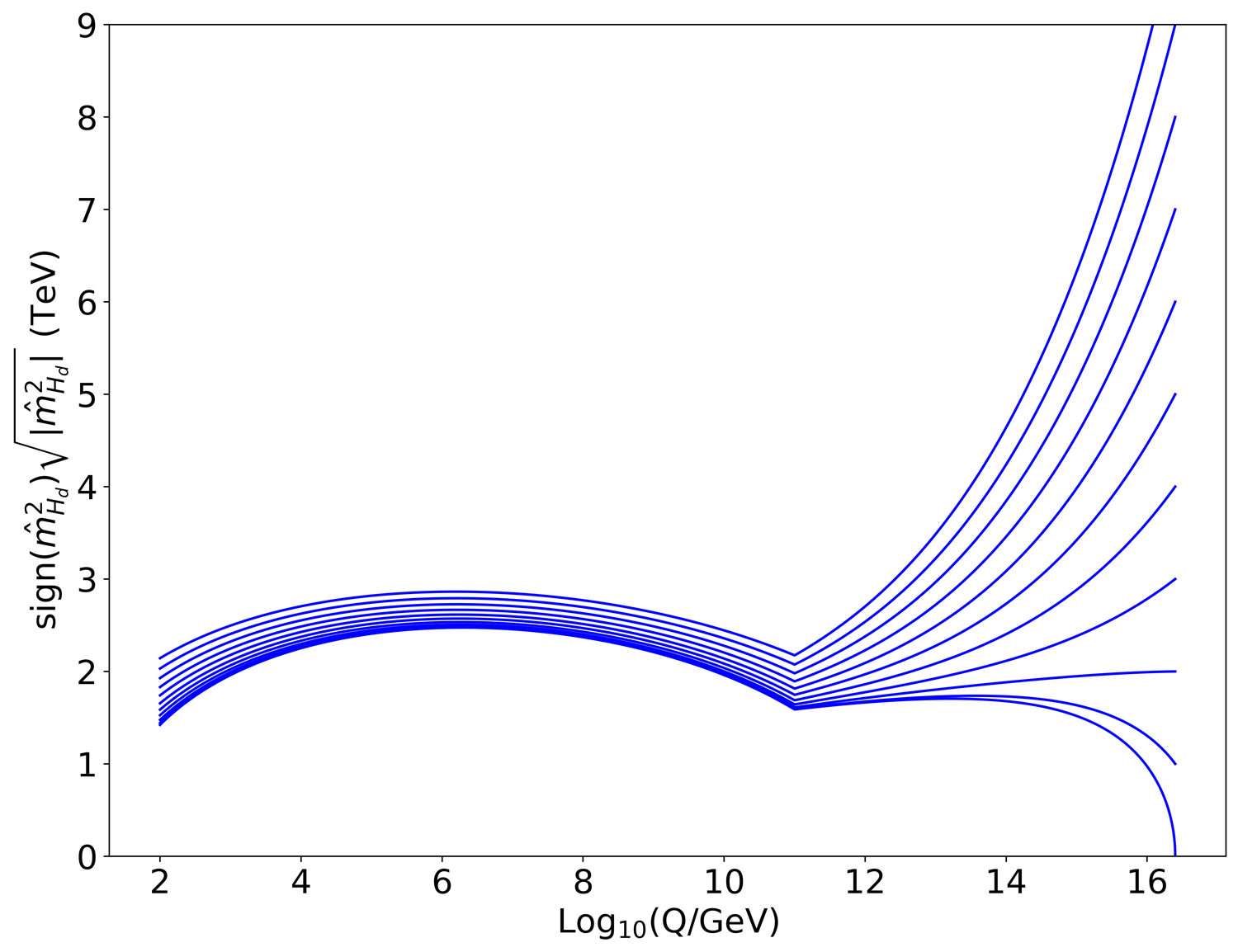}

			\caption{Running of {\it a}) $\hat{m}_{H_u}^2$, {\it b}) $m_{H_u}^2$ and $\mu$
                          and {\it c}) $\hat{m}_{H_d}^2$ from $Q=m_{GUT}$ to
				$Q=m_{weak}$ under the SPM scheme with $\Gamma =0.3$.
				}
		\label{fig:quasi1}
		\end{center}
	\end{figure}

	The approach to the quasifixed point values are shown in
	Figs. \ref{fig:quasi2} and \ref{fig:quasi1} for several $m_2^2$ cases
        (some of these results verify similar plots by SPM but are presented again
        here for the benefit of the reader\cite{Martin:2017vlf}).
        In Fig. \ref{fig:quasi2}, we show running of {\it a}) the bilinear soft term
          $B\equiv b/\mu$, {\it b}) running of $m_{\tu_R}$ and $M_3$ and {\it c})
          running of $m_{\te_R}$ and $M_1$. In all frames, we take $m_{1/2}=4.5$ TeV, $M_{int}=10^{11}$ GeV
          with all matter and Higgs soft masses (and bilinear $b$) set to $m_0$ ($m_0^2$).
          A different curve is plotted for different values of $m_0$ ranging from 1 GeV to 10 TeV.
          The values of $A_0$ and $\mu$ are solved for in each individual curve.
          Blue curves are for $\mu >0$ while purple curves are for $\mu <0$. From frame {\it a}),
          we see the quasi-fixed point for $b/\mu$ is largely small but negative. From frame {\it b}),
          we see that the typical squark mass approaches (in this case) a quasi-fixed point value
          around $\sim 2$ TeV rather than zero as in the PRS scheme; this helps avoid CCB minima
          in the SPM scheme.
          Squark soft masses are then pulled to large values at $m_{weak}$ due to the large value of
          $M_3$ which is chosen.
          From frame {\it c}), we see that typical slepton masses approach $\sim 1$ TeV at $M_{int}$
          in SPM rather than zero as in PRS. This helps avoid slepton LSP issues in the SPM scheme.

        In Fig. \ref{fig:quasi1}, we show in frame {\it a}) the running of
        $sign (\hat{m}_{H_u}^2)\sqrt{|\hat{m}_{H_u}^2|}$.
          This also runs to a quasi-fixed point,
            in this case around $\sim 2$ TeV. In frame {\it b}), we show the individual running
            of $m_{H_u}^2$ and $\mu $. While $\mu$ runs nearly flat as expected, $m_{H_u}^2$ runs
            at first to large negative values, and then below $M_{int}$ runs nearly flat.
            This example may assuage concerns that the running of $m_{H_u}^2$ below $M_{int}$ may destroy
            its correlation with $\mu$ so that the two terms regain some measure of independence:
            this doesn't seem to happen. In frame {\it c}), the running of $\hat{m}_{H_d}^2$ is shown
            and again it runs to a quasi-fixed point around 2 TeV.

\begin{figure}[!htbp]
		\begin{center}
			\includegraphics[height=0.4\textheight]{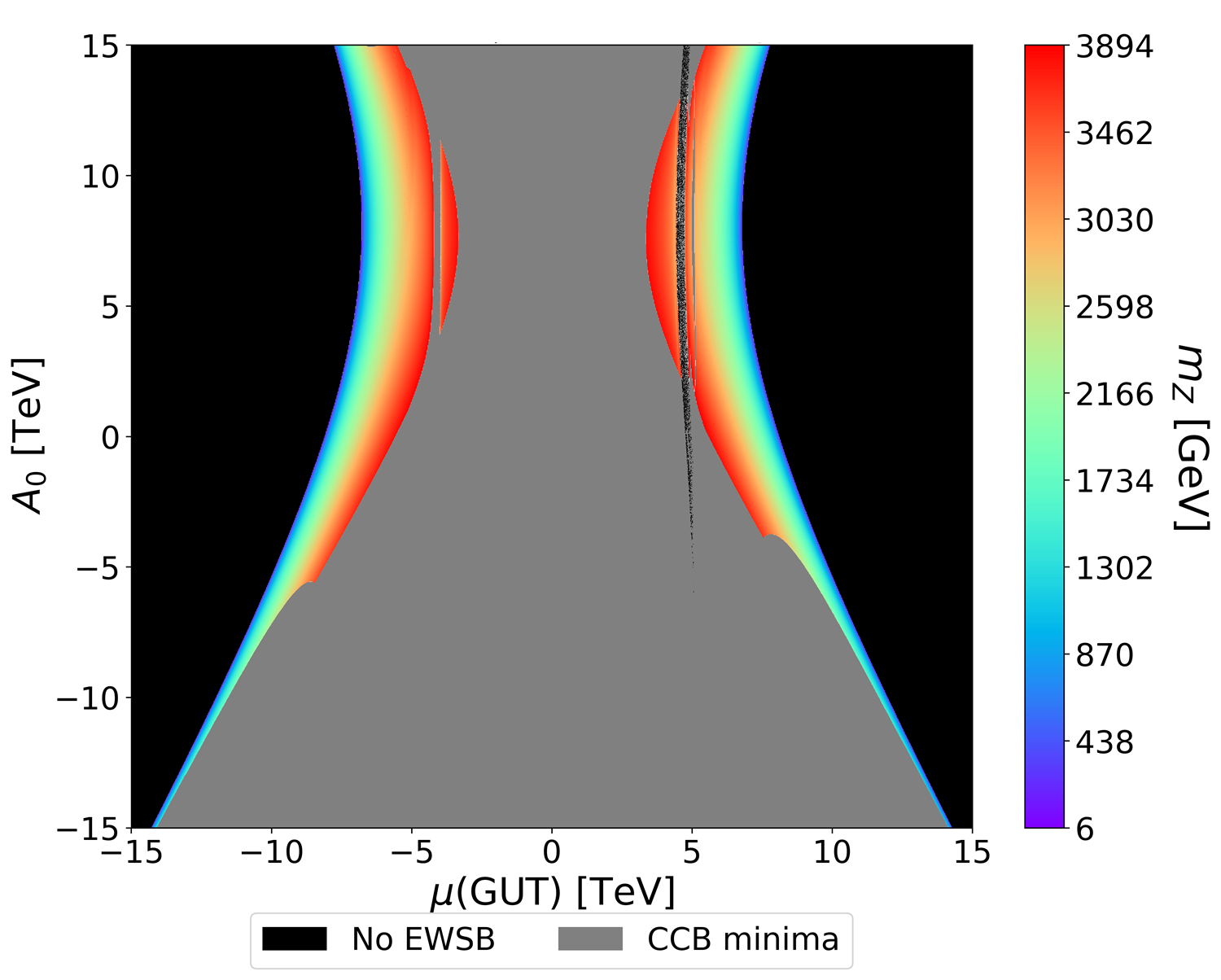}
			\caption{Color-coded regions of the derived value of $m_Z$ in the
				$\mu$ vs. $A_0$ plane for $\tan\beta =15$ with $m_0=10$ TeV and $m_{1/2}=4.5$ TeV.
				\label{fig:mZ}}
		\end{center}
	\end{figure}

	In the SPM scheme, the $m_2^2$ values approach (but do not exactly meet)
	their quasifixed point values at $Q= M_{int}$, so that the boundary conditions
	at $Q=M_{int}$ are no longer fixed. Thus, to generate a workable model,
	we must expand the parameter space from the PRS scheme.
	For SPM, therefore, one must reintroduce the various $m_2^2$ boundary conditions at $Q=m_*$,
        and we will take
	\be
	m_0,\ m_{1/2},\ A_0,\ \mu,\ b=B\mu .
	\ee
	After checking for appropriate EWSB, and then employing the EWSB minimization
	conditions, one can again solve for the derived value of $m_Z$.
	This is shown in Fig. \ref{fig:mZ} where we show color-coded regions of $m_Z$
	in the $A_0$ vs. $\mu (GUT)$ plane for $m_0=10$ TeV, $m_{1/2}=4.5$ TeV
	and $\tan\beta =15$.
        From the plot, one sees that there is no unique solution for $m_Z\simeq 91.2$ GeV
        but rather two disconnected regions depending
	on the sign of $\mu$, with a different $\mu$ value being obtained for each choice of $A_0$.

\subsection{Case with UGMs}
        
\begin{figure}[!htbp]
		\begin{center}
			\includegraphics[height=0.4\textheight]{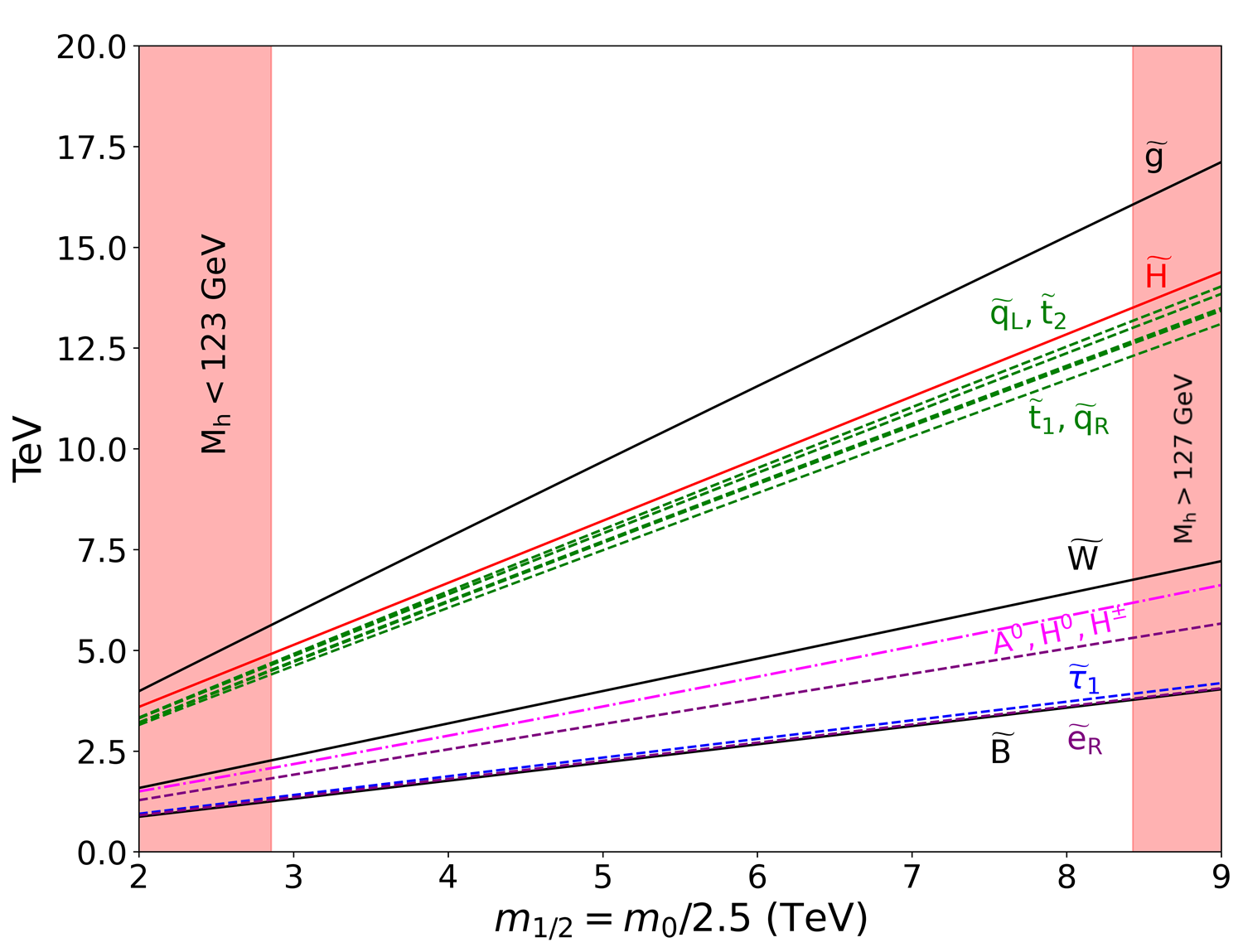}
			\caption{Sparticle masses vs. $m_{1/2}$ in the SPM UGM case
                          with $m_{1/2}=m_0 /2.5$.
				\label{fig:mass1}}
		\end{center}
	\end{figure}

	In the SPM paper, Martin has plotted out sample spectra for two cases,
	one with unified gaugino masses (UGM) and one with non-unified gaugino masses
	(NUGM). For the case of UGM, he shows sparticle mass spectra
	vs. $m_{1/2}$ for $m_0=m_{1/2}$ and also for $m_0=2.5 m_{1/2}$,
	with $\tan\beta =15$ and with $b=m_{1/2}^2$, where $A_0$ and $\mu$
	are solved for.
	For the $m_0=m_{1/2}$ case, he always finds a right selectron as the LSP
	(as do we), so that either additional $R$-parity violating
	interactions or lighter DM particles (such as axino) are needed to
	avoid charged stable relics from the early universe.
	In the case of $m_0\agt 2.5 m_{1/2}$, then the bino can become LSP.
	In Fig. \ref{fig:mass1}, we reproduce these results for the case
	of $m_0=2.5 m_{1/2}$. The region between the pink shaded
	boundaries has $123$ GeV $<m_h<127$ GeV (as computed here using
	FeynHiggs\cite{Hahn:2009zz}).
	Typically, in such cases with heavy sparticles in the
	multi-TeV range and a bino as the LSP, the thermally-produced
	neutralino relic density $\Omega_\chi h^2\gg 0.12$.
	However, from Fig. \ref{fig:mass1} we do see that since slepton
	masses are very nearly equal to $m(\text{bino})$, then coannihilation is
	available to reproduce the measured DM relic density. 
	We also see that the higgsino mass $\simeq \mu$ is very large, varying from
	$\sim 5-10$ TeV over the range of $m_{1/2}$ shown. This would make the model
	very unnatural under the conservative $\Delta_{EW}$ measure.
	However, the point here is that a mechanism is now present to
	drive the combination $m_{H_u}^2+\mu^2$ to small values, thus
	potentially ameliorating the LHP.

\begin{figure}[!htbp]
		\begin{center}
			\includegraphics[height=0.35\textheight]{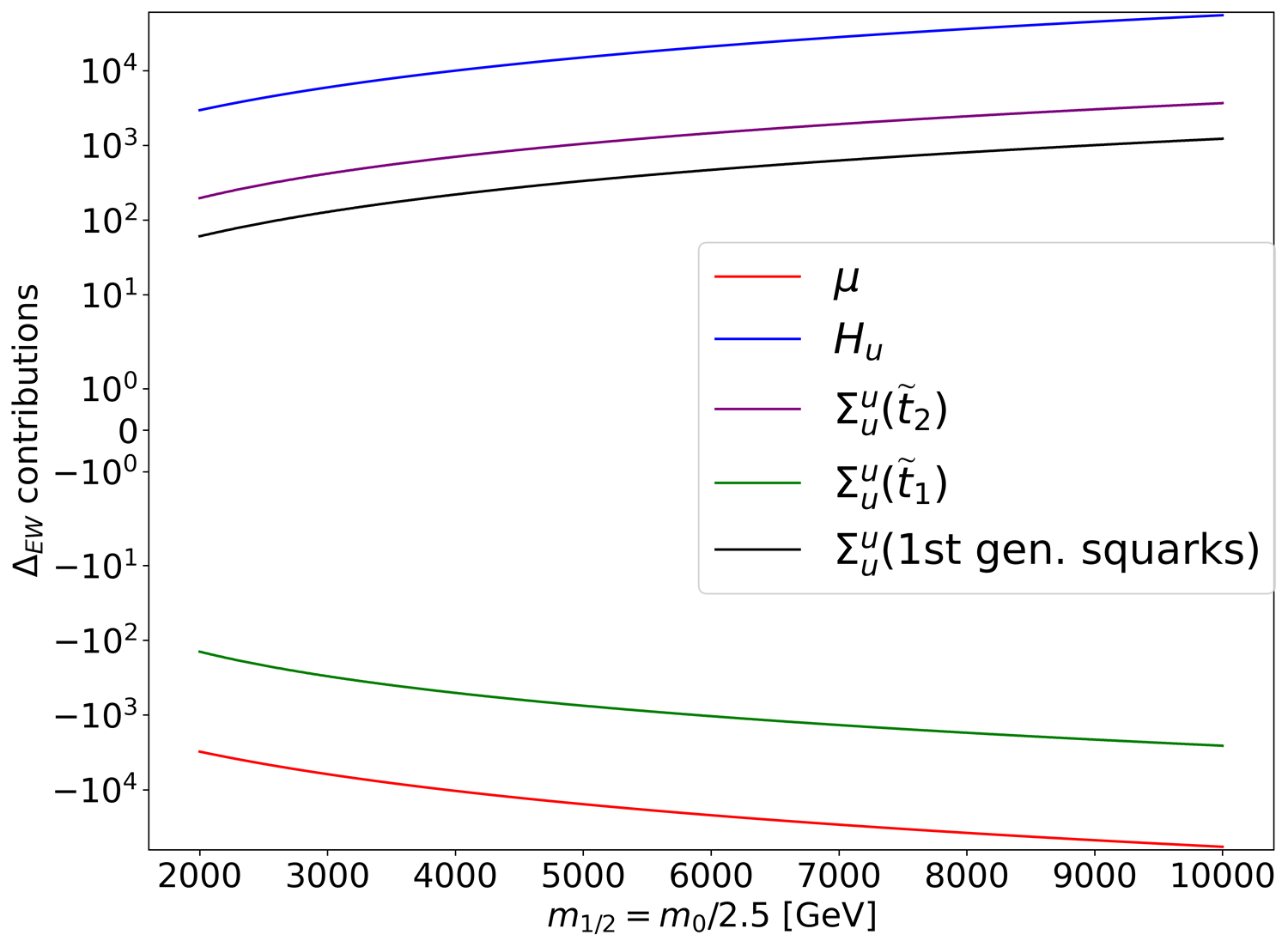}\\
			\includegraphics[height=0.35\textheight]{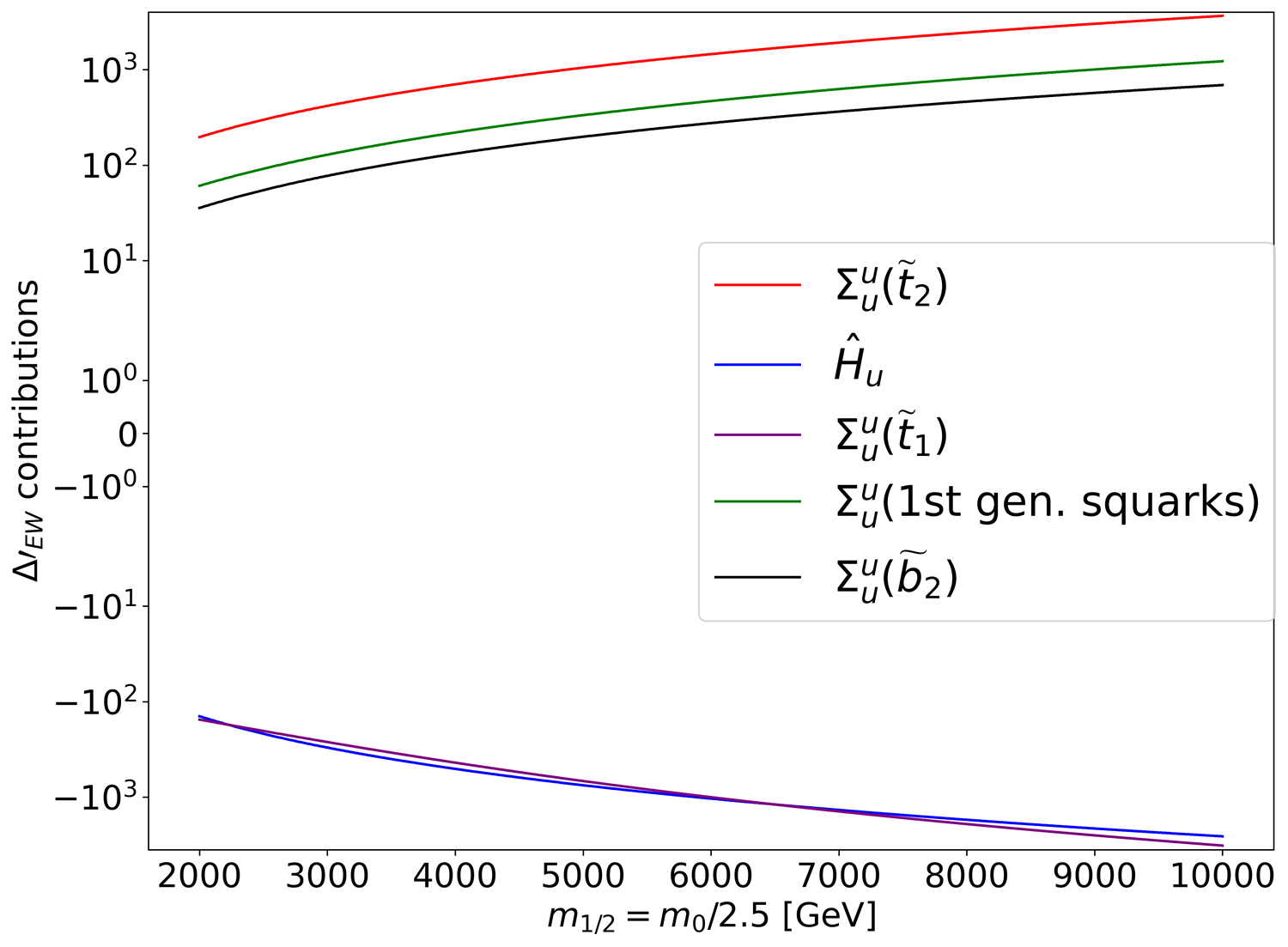}
			\caption{Top five signed contributions to {\it a}) $\Delta_{EW}$ and {\it b})
				$\Delta_{EW}^\prime$ for the UGM spectra with $m_0=2.5 m_{1/2}$.
				\label{fig:dew_dewp}}
		\end{center}
	\end{figure}

	Since we have now arrived at acceptable spectra for the case of
	scalar sequestering in the SPM scheme, we next want to check whether it
	really solves the LHP.
	In Fig. \ref{fig:dew_dewp}, we compute in frame {\it a}) the
	top five signed contributions to the naturalness measure $\Delta_{EW}$.
	The largest contributions come from $\mu$ and $m_{H_u}(weak)$,
	which are seen here as the blue and red curves. These lie in the $\sim 10^4$
	range in magnitude, making the model highly finetuned under $\Delta_{EW}$.
	In frame {\it b}), we define a revised finetuning measure $\Delta^\prime_{EW}$,
	which is the same as $\Delta_{EW}$ except that now $m_{H_u}^2+\mu^2$
	and $m_{H_d}^2+\mu^2$ are combined into single entities since they are now
	{\it dependent} (due to the CFT running above $Q= M_{int}$).
	In frame {\it b}), we see the top five contributions to
	$\Delta^\prime_{EW}$. In this case, the $\Sigma_u^u(\tst_{1,2})$ terms and
	$\hat{m}_{H_u}^2$ terms are largest, typically of order $\sim 10^3$.
	Thus, we find that although the SPM scheme in the UGM case has reduced
	finetuning, it is still found to be highly finetuned,
        mainly due to the large lightly-mixed
	top-squark masses contributing to the radiative corrections
	$\Sigma_u^u(\tst_{1,2})$.

	\subsection{Case with NUGM}
	
\begin{figure}[!htbp]
		\begin{center}
			\includegraphics[height=0.4\textheight]{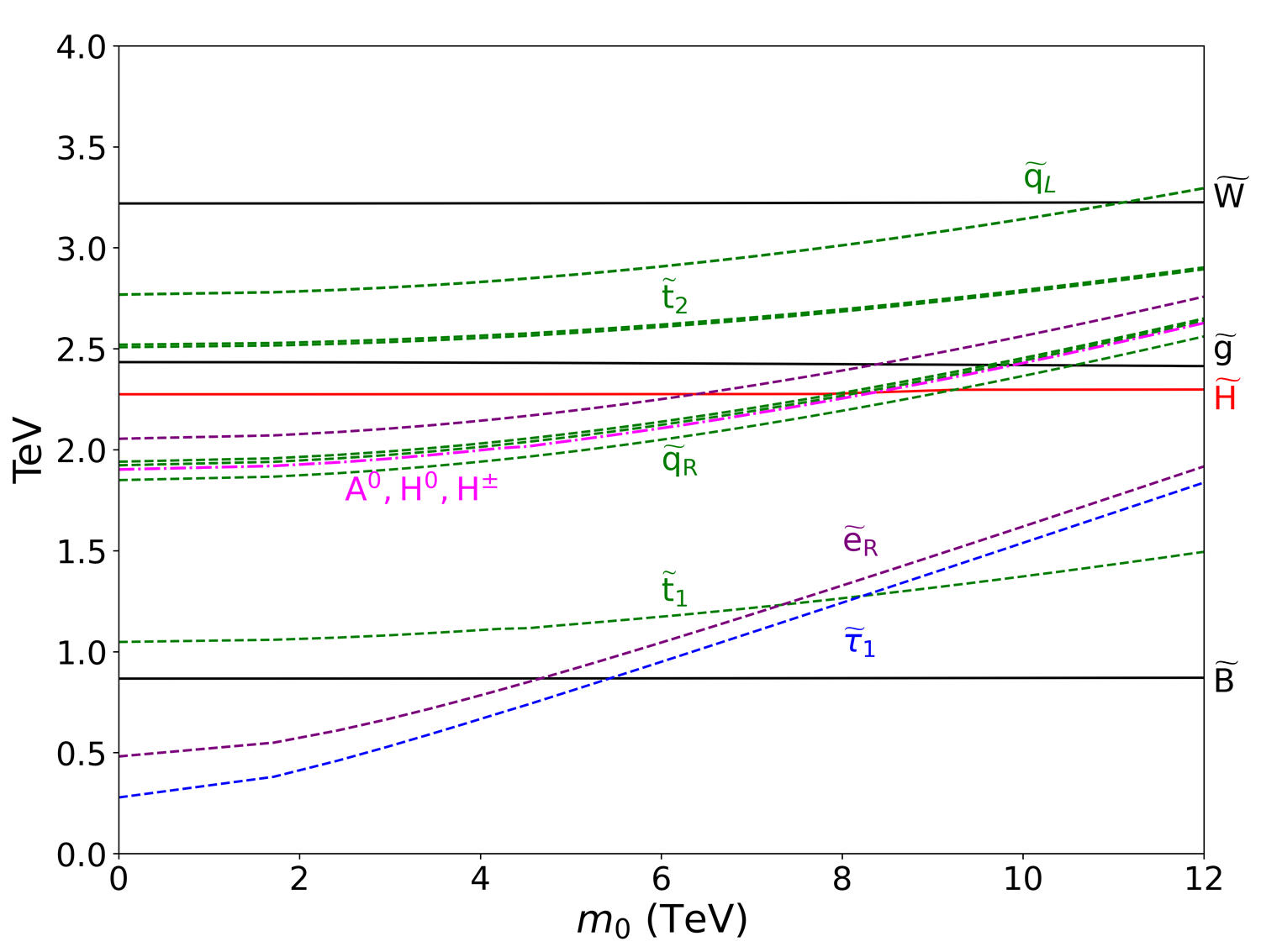}
			\caption{Sparticle and Higgs masses vs. $m_{0}$ in the SPM scheme
				with NUGM where $M_1=2$ TeV, $M_2=4$ TeV, and $M_3=1.2$ TeV,
				with $b=(2\ {\rm TeV})^2$, $\tan\beta =15$
                                and $\mu >0$.
				\label{fig:mass_nugm}}
		\end{center}
	\end{figure}

	Along with the UGM case, SPM also considers the case with NUGMs.
	This case is motivated by obtaining a large stop mixing element
	$A_t$ which can enhance $m_h\to 125$ GeV via maximal stop
	mixing rather than too large of stop masses. This can be achieved
	with $M_3\ll M_2$ while adjusting $M_1$ so that the bino remains
	as the LSP. In Fig. \ref{fig:mass_nugm}, we show the weak scale
	sparticle mass spectra in the SPM scheme with NUGMs.
        We plot vs. $m_0$ where $M_3=1.2$ TeV, $M_{2}=4$ TeV,
        and $M_1=2$ TeV (all $M_i$ defined at $Q=m_{GUT}$).
	Our calculations match well with the results of SPM.
	From the plot, we see that for low $m_0$ we still get a slepton as the LSP
	(this time, it is the $\tau$-slepton $\ttau_1$). For higher values of
	$m_0$, then sfermion masses increase as expected and for
	$m_0\agt 6$ TeV one obtains $m_{\tell}\agt m(bino)$ and so we get a
	bino as the LSP. Also, with $M_3(m_{GUT})$ only 1.2 TeV, then squarks and
	sleptons are much lighter than in Fig. \ref{fig:mass1}.
	With large stop mixing, then $m_h\sim 125$ GeV with not-too-heavy of
	stops and a chance for naturalness.
	The higgsinos are heavy and lie near $|\mu|\simeq m_{\tilde{H}}\sim 2.3$ TeV.

\begin{figure}[!htbp]
		\begin{center}
			\includegraphics[height=0.35\textheight]{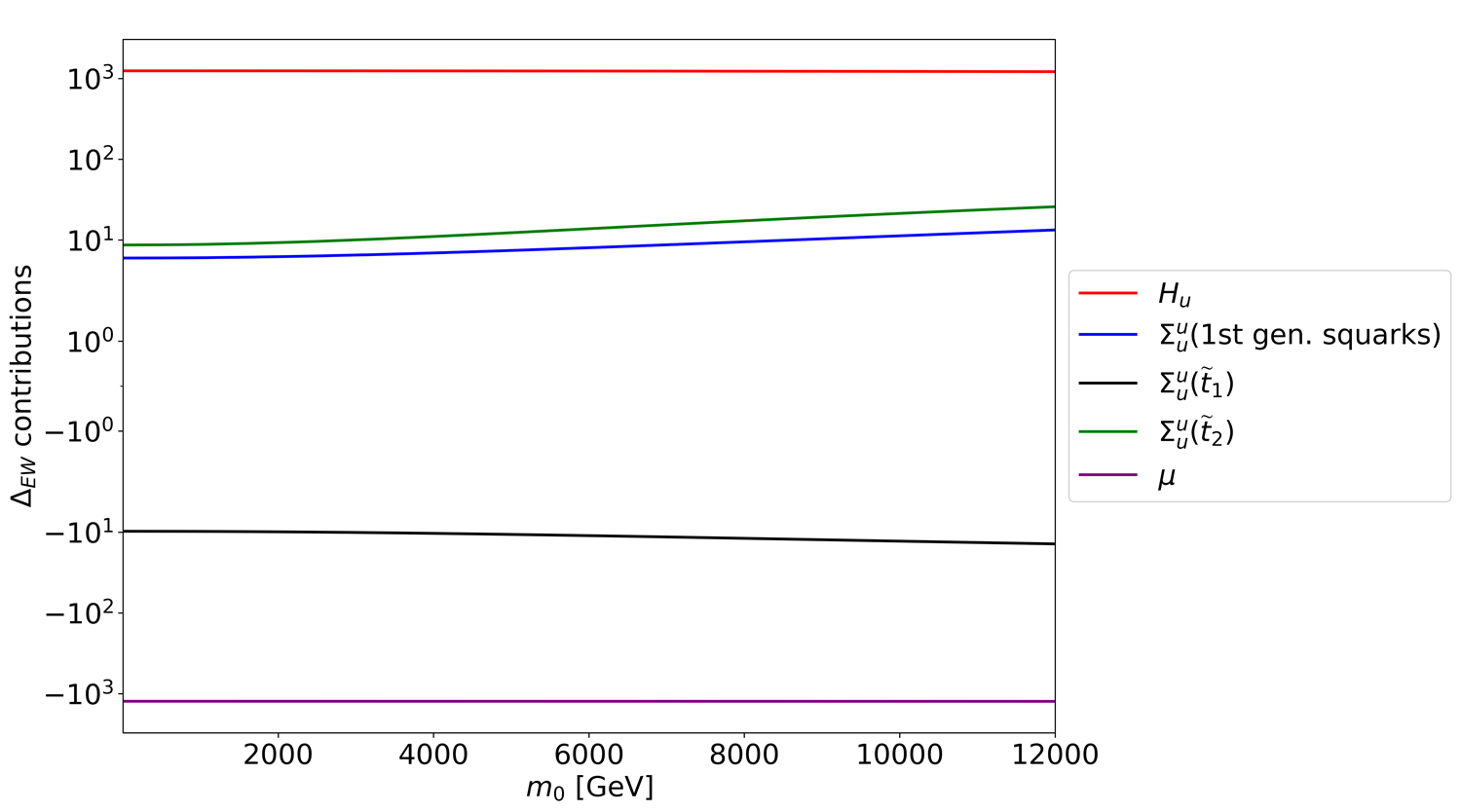}\\
			\includegraphics[height=0.35\textheight]{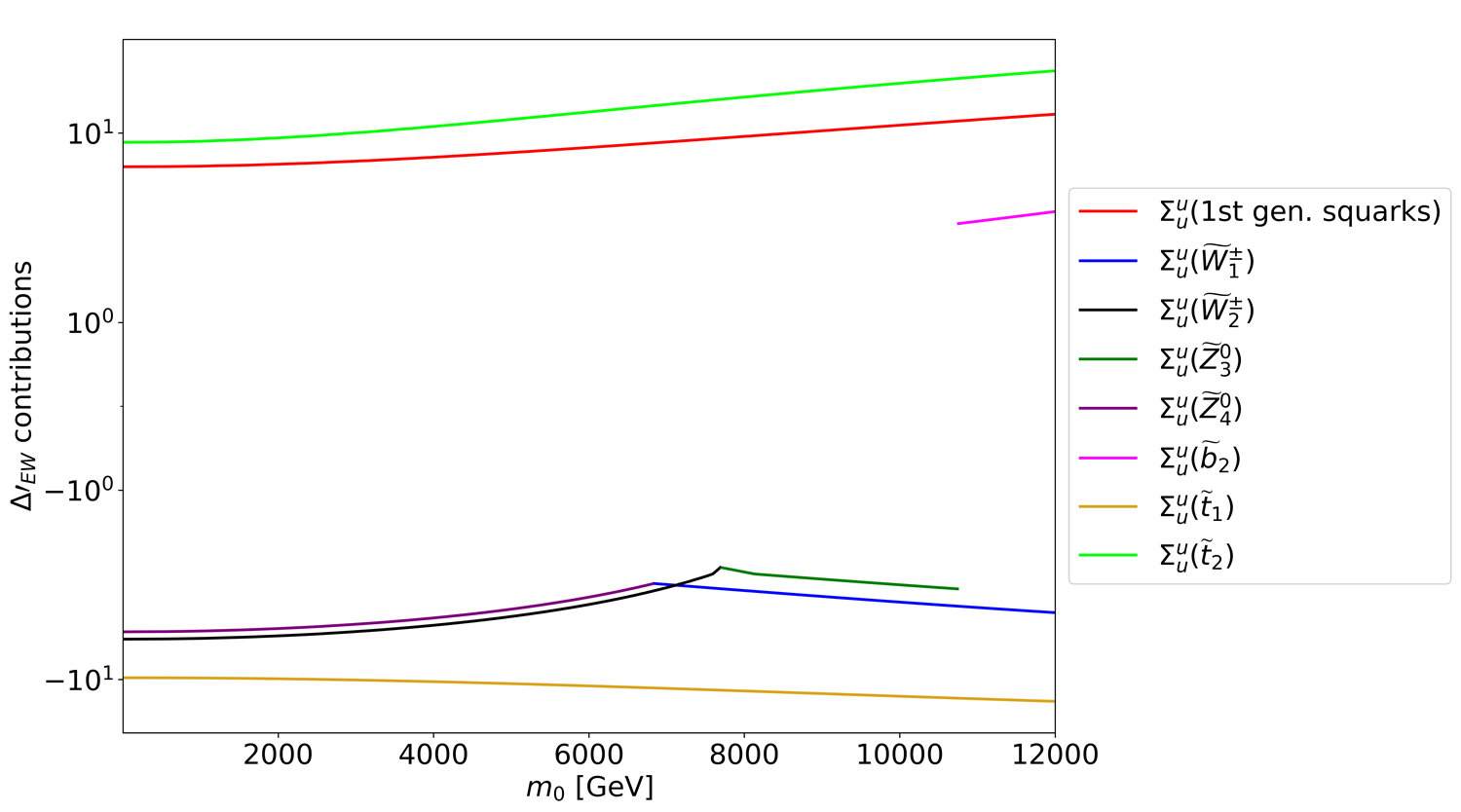}
			\caption{Top five signed contributions to {\it a}) $\Delta_{EW}$ and
				{\it b}) $\Delta^\prime_{EW}$ vs. $m_0$ for the SPM scheme with NUGM
				where $M_1=2$ TeV, $M_2=4$ TeV, and $M_3=1.2$ TeV,
				with $b=(2\ {\rm TeV})^2$ and $\tan\beta =15$.
				\label{fig:dew_dewp_nugm}}
		\end{center}
	\end{figure}

	In Fig. \ref{fig:dew_dewp_nugm} we compute the top five signed
	contributions to the finetuning measures
	{\it a}) $\Delta_{EW}$ and {\it b}) $\Delta^\prime_{EW}$ for the same
	parameters as in Fig. \ref{fig:mass_nugm}.
	From frame {\it a}), we see that the $m_{H_u}$ and $\mu$ contributions to
	$\Delta_{EW}$ are opposite sign but with absolute values $\sim 10^3$
	so that the spectra are finetuned under $\Delta_{EW}$.
	However, the SS of $m_{H_{u,d}}^2+\mu^2$ means these quantities are no longer
	independent and instead $\Delta^\prime_{EW}$ should be used.
	From frame {\it b}), we see the top five contributions to $\Delta^\prime_{EW}$
	are typically of order $\sim 10$: thus, this case of the SPM scheme with
	NUGMs seems natural even with higgsino masses of $\sim 2.3$ TeV.
	(The breaks in the curves of frame {\it b}) occur due to different
	contributions to Eq. \ref{eq:mzs} vying to be within the top five.)

	\section{Conclusions}
	\label{sec:conclude}
	
	Supersymmetric models with $m_h\sim 125$ GeV which obey LHC search constraints
	but have low $\Delta_{EW}$ qualify as permitting a SUSY solution to the GHP
	while avoiding the LHP: they are electroweak natural and are typified by
	the presence of light higgsinos with mass $m_{\tilde{H}}\alt 350$ GeV.
	Such light higgsinos are actively being searched for by ATLAS and CMS
	and indeed both experiments have some small excesses in the OSDLJMET signal
	channels. However, here we have investigated numerically the proposition
	whether hidden sector scalar sequestering can yield a solution to
	the LHP even with large $\mu\gg 350$ GeV. The key element is to
	assume a nearly superconformal hidden sector coupled to the visible sector
	which leads to driving scalar masses, the $B\mu$ term and the Higgs
	combinations $m_{H_{u,d}}^2+\mu^2$ $\rightarrow 0$ at some intermediate scale
	where the conformal symmetry is broken and the hidden sector is integrated out.
        A key ingredient is that the SUSY $\mu$ parameter is generatd via the GM mechanism.
	In such a situation, then HS dynamics pushes $m_{H_{u,d}}^2\sim -\mu^2$ so
	these quantities are no longer independent. Then a revised naturalness
	measure $\Delta^\prime_{EW}$ must be used where the now dependent
	terms $m_{H_{u,d}}^2$ and $\mu^2$ are combined.
	
	We investigated two schemes.
        \bi
        \item The strong scalar sequestering where scalar masses,
        $B\mu$ and Higgs combinations are driven to (nearly) zero at the intermediate
	scale (PRS scheme). This scheme has trouble developing appropriate
	electroweak symmetry breaking; and when it does, it develops a slepton
	as the LSP.
        Via thorough scans of parameter space, with both UGMs and NUGMs,
        we find no viable spectra with appropriate EWSB but 
	without a slepton as the LSP. Some of the problematic slepton LSPs can be avoided by
	introducing new sparticle or interactions which allow one to evade cosmological
	constraints on charged stable relics from the Big Bang.
	
	\item The second scheme labeled as SPM introduces moderate scalar sequestering
	  with non-negligible MSSM running so that scalar masses, $B\mu$ and Higgs
          combinations run to quasi-fixed points rather than zero.
	In the SPM scheme for UGMs, spectra with neutral stable LSPs can be found,
	and the finetuning can be reduced, but not eliminated. For the case of
	NUGMs, then $m_h\sim 125$ GeV can be found with sparticles beyond
	LHC bounds {\it and} with low EW finetuning as found from the
	$\Delta^\prime_{EW}$ measure. The latter spectra can have rather heavy higgsinos
	since the sequestering leads to non-independent Higgs soft terms and the
	$\mu$ term. Determining the consequences of the viable SPM scheme with NUGMs
        at colliding beam experiments is a topic for future work.
        \ei
	
	\section*{Acknowledgements} 
	
	This material is based upon work supported by the U.S. Department of Energy, 
	Office of Science, Office of Basic Energy Sciences Energy Frontier Research 
	Centers program under Award Number DE-SC-0009956.
        VB gratefully acknowledges support from the William F. Vilas estate.

	
	
	
	\bibliography{seq.bib}

\begin{thebibliography}{10}
\expandafter\ifx\csname url\endcsname\relax
  \def\url#1{\texttt{#1}}\fi
\expandafter\ifx\csname urlprefix\endcsname\relax\def\urlprefix{URL }\fi
\expandafter\ifx\csname href\endcsname\relax
  \def\href#1#2{#2} \def\path#1{#1}\fi

\bibitem{Baer:2006rs}
H.~Baer, X.~Tata, {Weak scale supersymmetry: From superfields to scattering
  events}, Cambridge University Press, 2006.

\bibitem{Dreiner:2023yus}
H.~K. Dreiner, H.~E. Haber, S.~P. Martin, {From Spinors to Supersymmetry},
  Cambridge University Press, Cambridge, UK, 2023.
\newblock \href {https://doi.org/10.1017/9781139049740}
  {\path{doi:10.1017/9781139049740}}.

\bibitem{Canepa:2019hph}
A.~Canepa, {Searches for Supersymmetry at the Large Hadron Collider}, Rev.
  Phys. 4 (2019) 100033.
\newblock \href {https://doi.org/10.1016/j.revip.2019.100033}
  {\path{doi:10.1016/j.revip.2019.100033}}.

\bibitem{Dine:2015xga}
M.~Dine, {Naturalness Under Stress}, Ann. Rev. Nucl. Part. Sci. 65 (2015)
  43--62.
\newblock \href {http://arxiv.org/abs/1501.01035} {\path{arXiv:1501.01035}},
  \href {https://doi.org/10.1146/annurev-nucl-102014-022053}
  {\path{doi:10.1146/annurev-nucl-102014-022053}}.

\bibitem{Baer:2012cf}
H.~Baer, V.~Barger, P.~Huang, D.~Mickelson, A.~Mustafayev, X.~Tata, {Radiative
  natural supersymmetry: Reconciling electroweak fine-tuning and the Higgs
  boson mass}, Phys. Rev. D 87~(11) (2013) 115028.
\newblock \href {http://arxiv.org/abs/1212.2655} {\path{arXiv:1212.2655}},
  \href {https://doi.org/10.1103/PhysRevD.87.115028}
  {\path{doi:10.1103/PhysRevD.87.115028}}.

\bibitem{Baer:2021tta}
H.~Baer, V.~Barger, D.~Martinez, {Comparison of SUSY spectra generators for
  natural SUSY and string landscape predictions}, Eur. Phys. J. C 82~(2) (2022)
  172.
\newblock \href {http://arxiv.org/abs/2111.03096} {\path{arXiv:2111.03096}},
  \href {https://doi.org/10.1140/epjc/s10052-022-10141-2}
  {\path{doi:10.1140/epjc/s10052-022-10141-2}}.

\bibitem{Baer:2023cvi}
H.~Baer, V.~Barger, D.~Martinez, S.~Salam, {Practical naturalness and its
  implications for weak scale supersymmetry}, Phys. Rev. D 108~(3) (2023)
  035050.
\newblock \href {http://arxiv.org/abs/2305.16125} {\path{arXiv:2305.16125}},
  \href {https://doi.org/10.1103/PhysRevD.108.035050}
  {\path{doi:10.1103/PhysRevD.108.035050}}.

\bibitem{Baer:2012up}
H.~Baer, V.~Barger, P.~Huang, A.~Mustafayev, X.~Tata, {Radiative natural SUSY
  with a 125 GeV Higgs boson}, Phys. Rev. Lett. 109 (2012) 161802.
\newblock \href {http://arxiv.org/abs/1207.3343} {\path{arXiv:1207.3343}},
  \href {https://doi.org/10.1103/PhysRevLett.109.161802}
  {\path{doi:10.1103/PhysRevLett.109.161802}}.

\bibitem{Baer:2011ec}
H.~Baer, V.~Barger, P.~Huang, {Hidden SUSY at the LHC: the light higgsino-world
  scenario and the role of a lepton collider}, JHEP 11 (2011) 031.
\newblock \href {http://arxiv.org/abs/1107.5581} {\path{arXiv:1107.5581}},
  \href {https://doi.org/10.1007/JHEP11(2011)031}
  {\path{doi:10.1007/JHEP11(2011)031}}.

\bibitem{Baer:2020sgm}
H.~Baer, V.~Barger, S.~Salam, D.~Sengupta, X.~Tata, {The LHC higgsino discovery
  plane for present and future SUSY searches}, Phys. Lett. B 810 (2020) 135777.
\newblock \href {http://arxiv.org/abs/2007.09252} {\path{arXiv:2007.09252}},
  \href {https://doi.org/10.1016/j.physletb.2020.135777}
  {\path{doi:10.1016/j.physletb.2020.135777}}.

\bibitem{Baer:2021srt}
H.~Baer, V.~Barger, D.~Sengupta, X.~Tata, {New angular and other cuts to
  improve the Higgsino signal at the LHC}, Phys. Rev. D 105~(9) (2022) 095017.
\newblock \href {http://arxiv.org/abs/2109.14030} {\path{arXiv:2109.14030}},
  \href {https://doi.org/10.1103/PhysRevD.105.095017}
  {\path{doi:10.1103/PhysRevD.105.095017}}.

\bibitem{ATLAS:2019lng}
G.~Aad, et~al., {Searches for electroweak production of supersymmetric
  particles with compressed mass spectra in $\sqrt{s}=$ 13 TeV $pp$ collisions
  with the ATLAS detector}, Phys. Rev. D 101~(5) (2020) 052005.
\newblock \href {http://arxiv.org/abs/1911.12606} {\path{arXiv:1911.12606}},
  \href {https://doi.org/10.1103/PhysRevD.101.052005}
  {\path{doi:10.1103/PhysRevD.101.052005}}.

\bibitem{CMS:2021edw}
A.~Tumasyan, et~al., {Search for supersymmetry in final states with two or
  three soft leptons and missing transverse momentum in proton-proton
  collisions at $ \sqrt{s} $ = 13 TeV}, JHEP 04 (2022) 091.
\newblock \href {http://arxiv.org/abs/2111.06296} {\path{arXiv:2111.06296}},
  \href {https://doi.org/10.1007/JHEP04(2022)091}
  {\path{doi:10.1007/JHEP04(2022)091}}.

\bibitem{Han:2014kaa}
Z.~Han, G.~D. Kribs, A.~Martin, A.~Menon, {Hunting quasidegenerate Higgsinos},
  Phys. Rev. D 89~(7) (2014) 075007.
\newblock \href {http://arxiv.org/abs/1401.1235} {\path{arXiv:1401.1235}},
  \href {https://doi.org/10.1103/PhysRevD.89.075007}
  {\path{doi:10.1103/PhysRevD.89.075007}}.

\bibitem{Baer:2014kya}
H.~Baer, A.~Mustafayev, X.~Tata, {Monojet plus soft dilepton signal from light
  higgsino pair production at LHC14}, Phys. Rev. D 90~(11) (2014) 115007.
\newblock \href {http://arxiv.org/abs/1409.7058} {\path{arXiv:1409.7058}},
  \href {https://doi.org/10.1103/PhysRevD.90.115007}
  {\path{doi:10.1103/PhysRevD.90.115007}}.

\bibitem{Murayama:2007ge}
H.~Murayama, Y.~Nomura, D.~Poland, {More visible effects of the hidden sector},
  Phys. Rev. D 77 (2008) 015005.
\newblock \href {http://arxiv.org/abs/0709.0775} {\path{arXiv:0709.0775}},
  \href {https://doi.org/10.1103/PhysRevD.77.015005}
  {\path{doi:10.1103/PhysRevD.77.015005}}.

\bibitem{Perez:2008ng}
G.~Perez, T.~S. Roy, M.~Schmaltz, {Phenomenology of SUSY with scalar
  sequestering}, Phys. Rev. D 79 (2009) 095016.
\newblock \href {http://arxiv.org/abs/0811.3206} {\path{arXiv:0811.3206}},
  \href {https://doi.org/10.1103/PhysRevD.79.095016}
  {\path{doi:10.1103/PhysRevD.79.095016}}.

\bibitem{Kim:2009sy}
H.~D. Kim, J.-H. Kim, {Higgs Phenomenology of Scalar Sequestering}, JHEP 05
  (2009) 040.
\newblock \href {http://arxiv.org/abs/0903.0025} {\path{arXiv:0903.0025}},
  \href {https://doi.org/10.1088/1126-6708/2009/05/040}
  {\path{doi:10.1088/1126-6708/2009/05/040}}.

\bibitem{Martin:2017vlf}
S.~P. Martin, {Quasifixed points from scalar sequestering and the little
  hierarchy problem in supersymmetry}, Phys. Rev. D 97~(3) (2018) 035006.
\newblock \href {http://arxiv.org/abs/1712.05806} {\path{arXiv:1712.05806}},
  \href {https://doi.org/10.1103/PhysRevD.97.035006}
  {\path{doi:10.1103/PhysRevD.97.035006}}.

\bibitem{Luty:2001zv}
M.~Luty, R.~Sundrum, {Anomaly mediated supersymmetry breaking in
  four-dimensions, naturally}, Phys. Rev. D 67 (2000) 045007.
\newblock \href {http://arxiv.org/abs/hep-th/0111231}
  {\path{arXiv:hep-th/0111231}}, \href
  {https://doi.org/10.1103/PhysRevD.67.045007}
  {\path{doi:10.1103/PhysRevD.67.045007}}.

\bibitem{Randall:1998uk}
L.~Randall, R.~Sundrum, {Out of this world supersymmetry breaking}, Nucl. Phys.
  B 557 (1999) 79--118.
\newblock \href {http://arxiv.org/abs/hep-th/9810155}
  {\path{arXiv:hep-th/9810155}}, \href
  {https://doi.org/10.1016/S0550-3213(99)00359-4}
  {\path{doi:10.1016/S0550-3213(99)00359-4}}.

\bibitem{Giudice:1998xp}
G.~F. Giudice, M.~A. Luty, H.~Murayama, R.~Rattazzi, {Gaugino mass without
  singlets}, JHEP 12 (1998) 027.
\newblock \href {http://arxiv.org/abs/hep-ph/9810442}
  {\path{arXiv:hep-ph/9810442}}, \href
  {https://doi.org/10.1088/1126-6708/1998/12/027}
  {\path{doi:10.1088/1126-6708/1998/12/027}}.

\bibitem{Anisimov:2001zz}
A.~Anisimov, M.~Dine, M.~Graesser, S.~D. Thomas, {Brane world SUSY breaking},
  Phys. Rev. D 65 (2002) 105011.
\newblock \href {http://arxiv.org/abs/hep-th/0111235}
  {\path{arXiv:hep-th/0111235}}, \href
  {https://doi.org/10.1103/PhysRevD.65.105011}
  {\path{doi:10.1103/PhysRevD.65.105011}}.

\bibitem{Dine:2004dv}
M.~Dine, P.~J. Fox, E.~Gorbatov, Y.~Shadmi, Y.~Shirman, S.~D. Thomas, {Visible
  effects of the hidden sector}, Phys. Rev. D 70 (2004) 045023.
\newblock \href {http://arxiv.org/abs/hep-ph/0405159}
  {\path{arXiv:hep-ph/0405159}}, \href
  {https://doi.org/10.1103/PhysRevD.70.045023}
  {\path{doi:10.1103/PhysRevD.70.045023}}.

\bibitem{Craig:2009rk}
N.~J. Craig, D.~Green, {On the Phenomenology of Strongly Coupled Hidden
  Sectors}, JHEP 09 (2009) 113.
\newblock \href {http://arxiv.org/abs/0905.4088} {\path{arXiv:0905.4088}},
  \href {https://doi.org/10.1088/1126-6708/2009/09/113}
  {\path{doi:10.1088/1126-6708/2009/09/113}}.

\bibitem{Giudice:1998bp}
G.~F. Giudice, R.~Rattazzi, {Theories with gauge mediated supersymmetry
  breaking}, Phys. Rept. 322 (1999) 419--499.
\newblock \href {http://arxiv.org/abs/hep-ph/9801271}
  {\path{arXiv:hep-ph/9801271}}, \href
  {https://doi.org/10.1016/S0370-1573(99)00042-3}
  {\path{doi:10.1016/S0370-1573(99)00042-3}}.

\bibitem{Roy:2007nz}
T.~S. Roy, M.~Schmaltz, {Hidden solution to the mu/Bmu problem in gauge
  mediation}, Phys. Rev. D 77 (2008) 095008.
\newblock \href {http://arxiv.org/abs/0708.3593} {\path{arXiv:0708.3593}},
  \href {https://doi.org/10.1103/PhysRevD.77.095008}
  {\path{doi:10.1103/PhysRevD.77.095008}}.

\bibitem{Giudice:1988yz}
G.~F. Giudice, A.~Masiero, {A Natural Solution to the mu Problem in
  Supergravity Theories}, Phys. Lett. B 206 (1988) 480--484.
\newblock \href {https://doi.org/10.1016/0370-2693(88)91613-9}
  {\path{doi:10.1016/0370-2693(88)91613-9}}.

\bibitem{Bae:2019dgg}
K.~J. Bae, H.~Baer, V.~Barger, D.~Sengupta, {Revisiting the SUSY $\mu$ problem
  and its solutions in the LHC era}, Phys. Rev. D 99~(11) (2019) 115027.
\newblock \href {http://arxiv.org/abs/1902.10748} {\path{arXiv:1902.10748}},
  \href {https://doi.org/10.1103/PhysRevD.99.115027}
  {\path{doi:10.1103/PhysRevD.99.115027}}.

\bibitem{Barger:1989rk}
V.~D. Barger, G.~F. Giudice, T.~Han, {Some New Aspects of Supersymmetry
  R-Parity Violating Interactions}, Phys. Rev. D 40 (1989) 2987.
\newblock \href {https://doi.org/10.1103/PhysRevD.40.2987}
  {\path{doi:10.1103/PhysRevD.40.2987}}.

\bibitem{Dreiner:1997uz}
H.~K. Dreiner, {An Introduction to explicit R-parity violation}, Adv. Ser.
  Direct. High Energy Phys. 21 (2010) 565--583.
\newblock \href {http://arxiv.org/abs/hep-ph/9707435}
  {\path{arXiv:hep-ph/9707435}}, \href
  {https://doi.org/10.1142/9789814307505_0017}
  {\path{doi:10.1142/9789814307505_0017}}.

\bibitem{Choi:2011yf}
K.-Y. Choi, L.~Covi, J.~E. Kim, L.~Roszkowski, {Axino Cold Dark Matter
  Revisited}, JHEP 04 (2012) 106.
\newblock \href {http://arxiv.org/abs/1108.2282} {\path{arXiv:1108.2282}},
  \href {https://doi.org/10.1007/JHEP04(2012)106}
  {\path{doi:10.1007/JHEP04(2012)106}}.

\bibitem{Baer:2014ica}
H.~Baer, V.~Barger, D.~Mickelson, M.~Padeffke-Kirkland, {SUSY models under
  siege: LHC constraints and electroweak fine-tuning}, Phys. Rev. D 89~(11)
  (2014) 115019.
\newblock \href {http://arxiv.org/abs/1404.2277} {\path{arXiv:1404.2277}},
  \href {https://doi.org/10.1103/PhysRevD.89.115019}
  {\path{doi:10.1103/PhysRevD.89.115019}}.

\bibitem{Carena:2002es}
M.~Carena, H.~E. Haber, {Higgs Boson Theory and Phenomenology}, Prog. Part.
  Nucl. Phys. 50 (2003) 63--152.
\newblock \href {http://arxiv.org/abs/hep-ph/0208209}
  {\path{arXiv:hep-ph/0208209}}, \href
  {https://doi.org/10.1016/S0146-6410(02)00177-1}
  {\path{doi:10.1016/S0146-6410(02)00177-1}}.

\bibitem{Martin:1994rge}
S.~P. Martin, M.~T. Vaughn, Two-loop renormalization group equations for soft
  supersymmetry-breaking couplings, Phys. Rev. D 50 (1994) 2282--2292.
\newblock \href {https://doi.org/10.1103/PhysRevD.50.2282}
  {\path{doi:10.1103/PhysRevD.50.2282}}.

\bibitem{Allanach:2002ss}
B.~Allanach, {SOFTSUSY: A program for calculating supersymmetric spectra},
  Computer Physics Communications 143~(3) (2002) 305–331.
\newblock \href {https://doi.org/10.1016/s0010-4655(01)00460-x}
  {\path{doi:10.1016/s0010-4655(01)00460-x}}.

\bibitem{Buckley:2015py}
A.~Buckley, {PySLHA: a Pythonic interface to SUSY Les Houches Accord data}
  (2015).
\newblock \href {http://arxiv.org/abs/1305.4194} {\path{arXiv:1305.4194}}.

\bibitem{Srednicki:1986vj}
M.~Srednicki, K.~A. Olive, J.~Silk, {High-Energy Neutrinos from the Sun and
  Cold Dark Matter}, Nucl. Phys. B 279 (1987) 804--823.
\newblock \href {https://doi.org/10.1016/0550-3213(87)90020-4}
  {\path{doi:10.1016/0550-3213(87)90020-4}}.

\bibitem{LZ:2022lsv}
J.~Aalbers, et~al., {First Dark Matter Search Results from the LUX-ZEPLIN (LZ)
  Experiment}, Phys. Rev. Lett. 131~(4) (2023) 041002.
\newblock \href {http://arxiv.org/abs/2207.03764} {\path{arXiv:2207.03764}},
  \href {https://doi.org/10.1103/PhysRevLett.131.041002}
  {\path{doi:10.1103/PhysRevLett.131.041002}}.

\bibitem{Poland:2011ey}
D.~Poland, D.~Simmons-Duffin, A.~Vichi, {Carving Out the Space of 4D CFTs},
  JHEP 05 (2012) 110.
\newblock \href {http://arxiv.org/abs/1109.5176} {\path{arXiv:1109.5176}},
  \href {https://doi.org/10.1007/JHEP05(2012)110}
  {\path{doi:10.1007/JHEP05(2012)110}}.

\bibitem{Poland:2015mta}
D.~Poland, A.~Stergiou, {Exploring the Minimal 4D $\mathcal{N}=1$ SCFT}, JHEP
  12 (2015) 121.
\newblock \href {http://arxiv.org/abs/1509.06368} {\path{arXiv:1509.06368}},
  \href {https://doi.org/10.1007/JHEP12(2015)121}
  {\path{doi:10.1007/JHEP12(2015)121}}.

\bibitem{Green:2012nqa}
D.~Green, D.~Shih, {Bounds on SCFTs from Conformal Perturbation Theory}, JHEP
  09 (2012) 026.
\newblock \href {http://arxiv.org/abs/1203.5129} {\path{arXiv:1203.5129}},
  \href {https://doi.org/10.1007/JHEP09(2012)026}
  {\path{doi:10.1007/JHEP09(2012)026}}.

\bibitem{Hahn:2009zz}
T.~Hahn, S.~Heinemeyer, W.~Hollik, H.~Rzehak, G.~Weiglein, {FeynHiggs: A
  program for the calculation of MSSM Higgs-boson observables - Version 2.6.5},
  Comput. Phys. Commun. 180 (2009) 1426--1427.
\newblock \href {https://doi.org/10.1016/j.cpc.2009.02.014}
  {\path{doi:10.1016/j.cpc.2009.02.014}}.

\end{thebibliography}
	\bibliographystyle{elsarticle-num}
	
\end{document}